%% file: medima-main.tex
\documentclass[a4paper,fleqn]{cas-dc}
\usepackage[authoryear]{natbib}
\pdfoutput=1 

\ExplSyntaxOn
\keys_set:nn { stm / mktitle } { nologo }
\ExplSyntaxOff

\usepackage{amssymb}
\usepackage{latexsym}

\usepackage{url}
\usepackage[dvipsnames]{xcolor}

\usepackage{booktabs}
\usepackage{graphicx}
\usepackage{hyperref}
\usepackage{caption}
\usepackage{subcaption}
\usepackage{amsmath}
\usepackage{comment}

\newcommand{\beginsupplement}{%
        \setcounter{table}{0}
        \renewcommand{\thetable}{S\arabic{table}}%
        \setcounter{figure}{0}
        \renewcommand{\thefigure}{S\arabic{figure}}%
     }

\begin{document}
\let\WriteBookmarks\relax
\def\floatpagepagefraction{1}
\def\textpagefraction{.001}



\shorttitle{Correlated Feature Aggregation by Region}

\shortauthors{Karin Stacke \textit{et~al.}}

\title[mode = title]{Correlated Feature Aggregation by Region Helps Distinguish Aggressive from Indolent Clear Cell Renal Cell Carcinoma Subtypes on CT}

\author[1,2,3]{Karin Stacke}[orcid=0000-0003-1066-3070]
\cormark[1]
\ead{karin.stacke@liu.se}

\author[3]{Indrani Bhattacharya}
\author[3]{Justin R. Tse}
\author[4]{James D. Brooks}
\author[3,4]{Geoffrey A. Sonn}
\fnmark[1]
\author[3]{Mirabela Rusu}
\fnmark[1]
\cormark[2]
\ead{mirabela.rusu@stanford.edu}
\fntext[fn2]{Equally contributed as senior authors.}

\address[1]{Division for Media and Information Technology, Department of Science and Technology, Linköping University, Linköping, Sweden}
\address[2]{Sectra, Linköping, Sweden}
\address[3]{Department of Radiology, Stanford University, Stanford, USA}
\address[4]{Department of Urology, Stanford University, Stanford, USA}


\cortext[cor1]{Corresponding author}
\cortext[cor2]{Principal corresponding author}

\begin{abstract}
Renal cell carcinoma (RCC) is a common cancer that varies in clinical behavior. Indolent RCC is often low-grade without necrosis and can be monitored without treatment. Aggressive RCC is often high-grade and can cause metastasis and death if not promptly detected and treated. Clear cell RCC is the most common subtype of RCC with a high degree of aggressive cases. While most kidney cancers are detected on CT scans, grading is based on histology from invasive biopsy or surgery. Determining aggressiveness on CT images would be an important clinical advance as it would facilitate risk stratification and treatment planning. The aim of this study is to use machine learning methods to identify radiology features that correlate with features on pathology to facilitate assessment of cancer aggressiveness on CT images instead of histology. 
This paper presents a novel automated method, Correlated Feature Aggregation By Region (CorrFABR), for classifying aggressiveness of clear cell RCC by leveraging correlations between radiology and corresponding unaligned pathology images. CorrFABR consists of three main steps: (1) \textit{Feature Aggregation} where region-level features are extracted from radiology and pathology images, (2) \textit{Fusion} where radiology features correlated with pathology features are learned on a region level, and (3) \textit{Prediction} where the learned correlated features are used to distinguish aggressive from indolent clear cell RCC using CT alone as input. Thus, during training, CorrFABR learns from both radiology and pathology images, but during inference, CorrFABR will distinguish aggressive from indolent clear cell RCC using CT alone, in the absence of pathology images. CorrFABR improved classification performance over radiology features alone, with an increase in binary classification F1-score from $0.68\pm0.04$ to $0.73\pm0.03$. This demonstrates the potential of incorporating pathology disease characteristics for improved classification of aggressiveness of clear cell RCC on CT images.
\end{abstract}

\begin{keywords}
Renal Cell Carcinoma  \sep Clear Cell Renal Cell Carcinoma \sep Prostate cancer \sep Machine learning  \sep Deep learning \sep Correlated Feature Learning \sep Radiology-Pathology fusion  \sep Multi-modal
\end{keywords}

\maketitle



\input{sections/0_introduction}
\input{sections/2a_data}
\input{sections/2b_method}

\input{sections/2c_experiments}

\input{sections/4_conclusion}

\section*{Declaration of Competing Interest}
Mirabela Rusu is a paid consultant for Roche, the conflict is unrelated to this research. Also, Mirabela Rusu has research grants from Phillips Healthcare. The authors declare that they have no known competing financial interests or personal relationships that could have appeared to influence the work reported in this paper.

\section*{Acknowledgements}
This work was supported by the Wallenberg AI and Autonomous Systems and Software Program (WASP-AI), the research environment ELLIIT, AIDA Vinnova grant 2017-02447, Linköping University Center for Industrial Information Technology (CENIIT), Departments of Radiology and Urology, Stanford University, GE Healthcare Blue Sky Award, National Institutes of Health, National Cancer Institute (U01CA196387, to J.D.B.), and the generous philanthropic support of our patients (G.S.). Research reported in this publication was supported by the National Cancer Institute of the National Institutes of Health under Award Number R37CA260346. The content is solely the responsibility of the authors and does not necessarily represent the official views of the National Institutes of Health.

\bibliographystyle{model2-names.bst}
\bibliography{medima-main}

\input{sections/supplementary_material}

\end{document}

%% file: sections/0_introduction.tex
\begin{figure*}
    \centering
    \begin{subfigure}[b]{0.42\linewidth}
        \centering
        \includegraphics[width=0.45\linewidth]{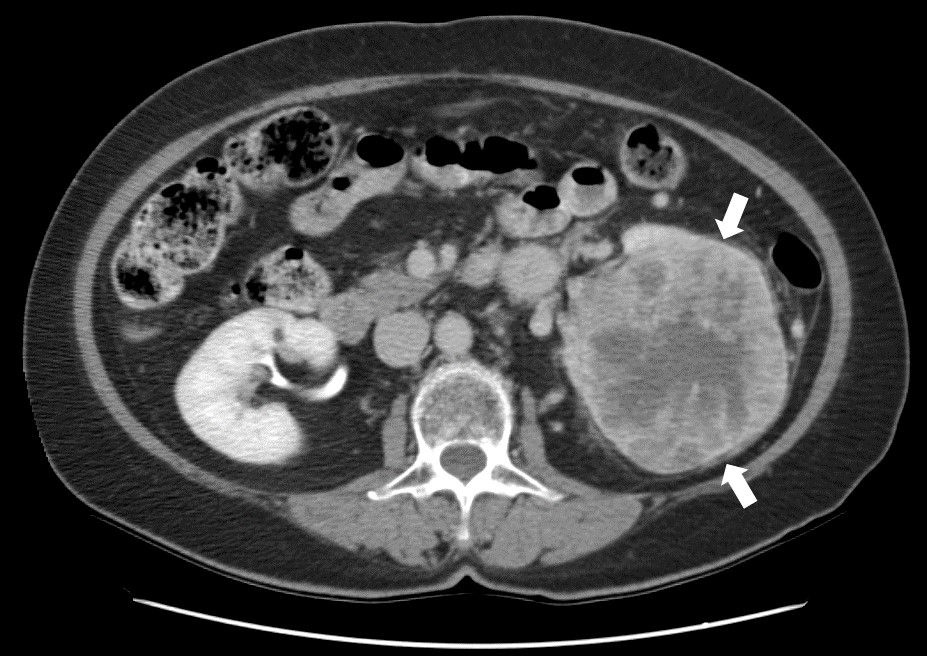}
        \includegraphics[width=0.41\linewidth]{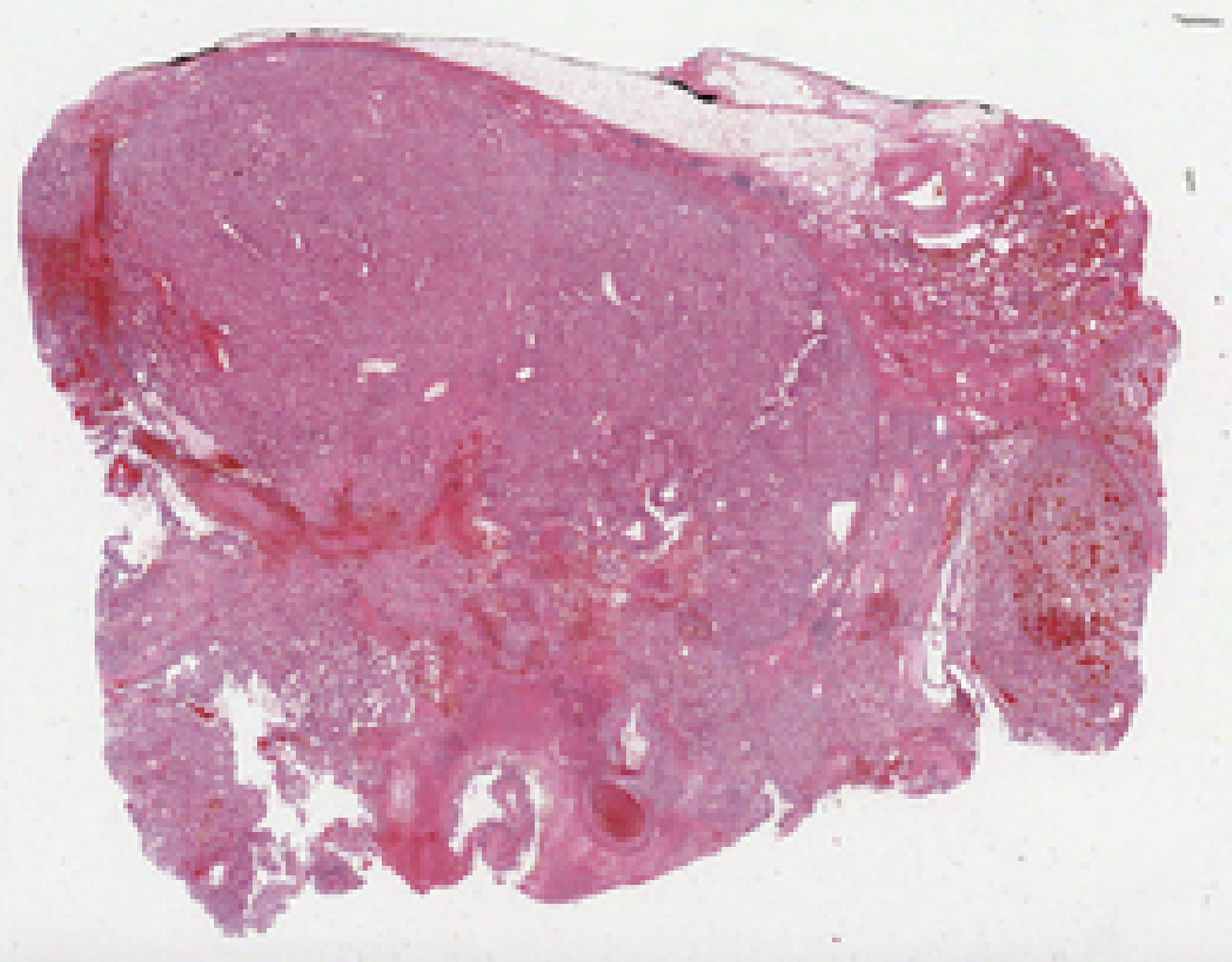}
        \caption{Kidney data}
        \label{fig:ex_kidney}
    \end{subfigure}
    \hfill
    \begin{subfigure}[b]{0.55\linewidth}
        \centering
        \includegraphics[width=0.31\textwidth]{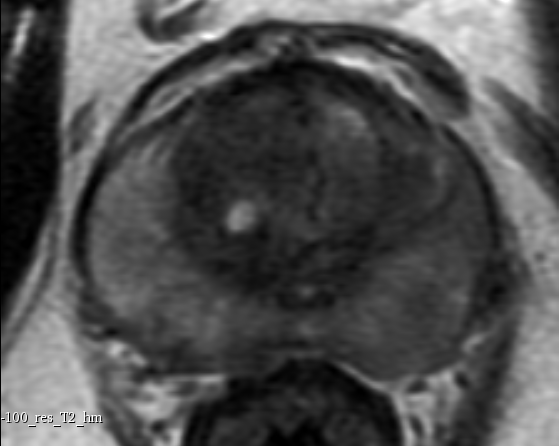}
        \includegraphics[width=0.31\textwidth]{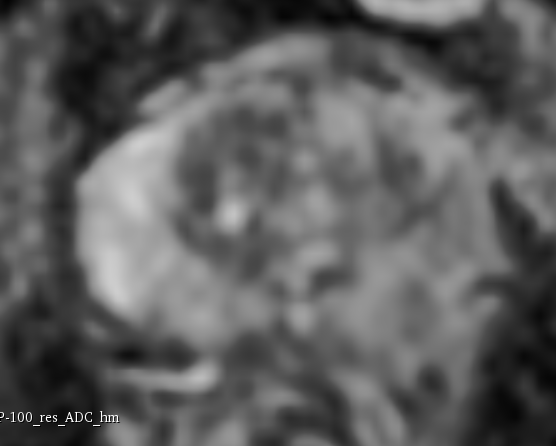}
        \includegraphics[width=0.31\textwidth]{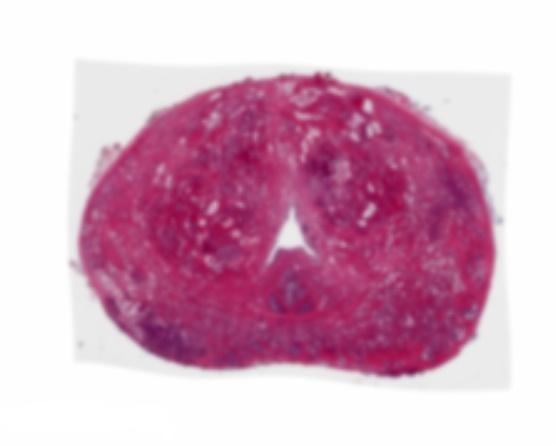}
        \caption{Prostate data}
        \label{fig:ex_prostate}
    \end{subfigure}
    \caption{Example images for kidney and prostate data.  (a) CT image (left) with an aggressive tumor highlighted with arrows, and the histopathology slice (right) showing a surgically resected tissue at an unknown location from the tumor. Please note the differences in resolution. (b) Spatially aligned T2w, ADC and histopathology images from the prostate cohort.}
    \label{fig:example_images}
\end{figure*}

\section{Introduction}

In 2021, there were nearly 500 000 new renal cell carcinoma (RCC) cases diagnosed, and 200 000 reported deaths world-wide~\citep{sung_2021}. 
Increased utilization of abdominal imaging has led to greatly increased detection of incidental renal masses \citep{capitanio2016renal, capitanio2019epidemiology}. Radiological imaging, such as computed tomography (CT), plays a vital role in risk assessment and treatment planning for these renal masses, by distinguishing benign masses from renal cell carcinoma, and by characterizing the aggressiveness of malignant masses~\citep{deleon_2017}. 

Clear cell RCC is the most common subtype ($\approx 75\%$), accountable for the majority of RCC deaths~\citep{hsieh_2017}. Although clear cell RCC is asymptomatic in early stages, 25\% to 30\% of patients have tumors that have metastasized at the time of diagnosis \citep{rasmussen2022artificial, padala_2020}. 
The prognosis of clear cell RCC depends on the aggressiveness of the mass, with higher tumor grade and the presence of necrosis being correlated with worse clinical outcomes \citep{minardi_2005, zhang_tumor_2018, moch_2016}. 

The aggressiveness of clear cell RCC using tumor grade and necrosis is assessed on pathology images of tissue collected via biopsy or surgery. These invasive procedures are often associated with adverse side effects of bleeding, pain, and infection. Making this assessment on non-invasive radiology images, based on features like tumor grade and necrosis, can help decide the optimum management strategy, and prevent overtreatment and adverse side effects in patients with indolent disease \citep{rasmussen2022artificial, campbell2017renal, rosales2010active}. 

While there are clear clinical benefits of classifying aggressiveness of clear cell RCC on radiology images, the classification is challenging due to the subtle and overlapping features of the two classes (indolent and aggressive)~\citep{morshid_2021, hotker_2016}. Correlating radiology and pathology features may help differentiate between the radiological appearance of aggressive and indolent clear cell RCC masses. Machine learning methods may help in learning these radiology-pathology correlations and in identifying pathology-informed radiology biomarkers for classification of clear cell RCC using radiology images alone.

Prior machine-learning based image-analysis methods for RCC focused on differentiating between malignant or benign masses~\citep{coy_2019, sun_2020, tanaka_2020}, classifying  histological subtypes~\citep{yan_2015, han_2019, zuo_2021, uhm_2021}, or distinguishing between low- and high-grade clear cell RCC~\citep{ding_2018, kocak_2019a, bektas_2019, lin_ct-based_2019, sun2019prediction, shu_clear_2019, nazari_noninvasive_2020, cui_2020, lin_2020, xv_2021, demirjian_2021}. In this study, we extend the definition of aggressiveness to include both tumor grade and presence of necrosis, as multivariate analysis gives better prognostic value~\citep{hotker_2016, tse_2021}. No prior study attempted to classify clear cell RCC as aggressive or indolent using machine learning, based on both tumor grade and necrosis observed on pathology images.

Only one study considered complementary histopathology images and genetic data in addition to radiology images for prediction of prognosis in clear cell RCC \citep{ning_2020}. The study showed that the multi-domain approach has the best performance, suggesting the benefit of multimodal data fusion. Yet, the requirement for all three modalities to be available as input data limits the clinical relevance of this approach, since both histological and genetic data may not be available in the pre-operative setting. 

Methods incorporating histopathology information during model training, while not requiring it during inference have been described for prostate cancer \citep{bhattacharya_2020, bhattacharya_2022}, but to our knowledge have not been developed for kidney cancer. The radiology-pathology fusion approach developed for the prostate can detect and differentiate between indolent and aggressive prostate cancer on Magnetic Resonance Imaging (MRI) \citep{bhattacharya_2022}. It is able to emphasize disease pathology characteristics on MR images by learning MRI features that are correlated with histopathology features on a per-pixel level. Thus, the method relies on registered (spatially aligned) MRI and histopathology images for creating these pathology-correlated radiology image biomarkers. The study showed that including MRI biomarkers improved the performance of prostate cancer detection over MR images alone.

Inspired by this prior study, we developed an approach to learn and use CT features in the kidney that correlate with histopathology image features for classification of clear cell RCC aggressiveness. However, application of the prior prostate approach to other cancer types is complicated by the dependence on spatially aligned radiology and pathology data, something usually lacking in datasets from other organ systems. In addition to the lack of registered radiology and pathology data, the clear cell RCC aggressiveness prediction task differs from the prostate task in the use of a different radiology imaging modality (CT vs. MRI), different organ (kidney vs. prostate) with different tissue characteristics in the histopathology images, and a different prediction challenge (classification vs. segmentation) (Figure~\ref{fig:example_images}). 

To enable accurate aggressiveness prediction of clear cell RCC on CT images, we present a novel radiology-pathology fusion approach, Correlated Feature Aggregation By Region (\textit{CorrFABR}), that identifies and uses radiology features that are correlated with disease pathology characteristics observed in resected tissue. Unlike the prior prostate study \citep{bhattacharya_2022}, CorrFABR does not require spatial alignment between radiology and pathology images. CorrFABR also includes a novel feature extraction module enabling feature aggregation from radiology and histopathology images at different resolutions.  

Our study has two parts. First, starting from a clinical dataset with existing spatial correspondences, CorrFABR is developed and trained on simulated prostate biopsy and surgery datasets \textit{without} pixel-wise spatial correspondences between the radiology and pathology images. Second, using public data without spatial correspondences, CorrFABR is trained and evaluated on cohorts from the 2021 Kidney Tumor Segmentation Challenge~\citep{heller_2021} and The Cancer Imaging Archive (TCIA)~\citep{clark_2013} to predict the aggressiveness of clear cell RCC. 

The main contributions of this paper, which are encompassed by the proposed CorrFABR method, can be summarized as:
\begin{enumerate}
    \item Development of radiology image biomarkers that correlate with pathology features without spatial alignment of the data through feature aggregation by region.
    \item Presentation of a novel approach for handling feature aggregation between radiology and pathology images with widely different resolutions.
    \item Demonstrate the application for indolent and aggressive characterization in clear cell RCC.
\end{enumerate}

The rest of the paper is organized as follows: the datasets used are presented in Section~\ref{sec:data}, followed by a description of the proposed method in Section~\ref{sec:method}. Section~\ref{sec:experiments} describes the experimental setup and presents results. In Section~\ref{sec:discussion}, the method and results are discussed.

%% file: sections/2a_data.tex
\section{Material and methods}

\begin{table*}[]
\centering
\caption{Dataset summary of the prostate and kidney cohorts. Available annotations of cancerous regions were either done manually by trained clinicians or automatically using pre-trained machine learning methods.}
\label{tab:cohorts}
\resizebox{\textwidth}{!}{%
\begin{tabular}{@{}lcccccccc@{}}
\toprule
& \multicolumn{5}{c}{Prostate} & \multicolumn{3}{c}{Kidney} \\ \cmidrule(lr){2-6} \cmidrule(lr){7-9}
& \multicolumn{3}{c}{$P_{cancer}$: Radical Prostatectomy} & \multicolumn{2}{c}{$P_{normal}$: Normal} & KiTS21 & \multicolumn{2}{c}{TCGA-KIRC}\\ \midrule
Number of Patients & \multicolumn{3}{c}{115} & \multicolumn{2}{c}{24} & 203 & \multicolumn{2}{c}{174}  \\
Sex (male/female) & \multicolumn{3}{c}{115 / 0} & \multicolumn{2}{c}{24 / 0} & 133 / 70 & \multicolumn{2}{c}{110 / 64} \\
Registered & \multicolumn{3}{c}{Yes} & \multicolumn{2}{c}{N/A} & N/A & \multicolumn{2}{c}{No} \\ 
Data  & \multicolumn{2}{c}{MRI} & Hist. & \multicolumn{2}{c}{MRI} & CT & CT & Hist. \\ \cmidrule(lr){2-4} \cmidrule(lr){5-6} \cmidrule(lr){7-7} \cmidrule(lr){8-9}
Sequence/Data Type & T2w & ADC & H\&E & T2w & ADC & Enhanced & Enhanced & H\&E \\
Number of slices (per volume) & 4--11 &  4--11 & 4--11 & 6--22 & 6--22 & 29--1059 & 32--306 & 1 \\

In-plane resolution (mm) & 0.27--0.94 & 0.78--1.50 & 0.008--0.016 & 0.35--0.43 & 0.78--1.25 & 0.59--1.04 & 0.50--0.98 & $[0.25-0.5] \cdot 10^{-3}$ \\

Distance between slices (mm) & 3.0--5.2 & 3.0--5.2 & 3.0–-5.2 & 3.0--4.2 & 3.0--4.2 & 0.5--5.0 & 1.25--7.5 & N/A \\ \midrule

Annotation type & Manual & Manual & Automatic & Manual & Manual & Manual & Automatic & Not available \\ \bottomrule
\end{tabular}%
}
\end{table*}

\subsection{Dataset}
\label{sec:data}
This work considers two different urologic cancer types: prostate cancer and renal (kidney) cancer. 

\subsubsection{Cohort Used for Simulated Prostate Data}

First, we used clinical prostate data (described below) to simulate lack of spatial alignment between radiology and pathology images. 
The clinical prostate dataset consists of two subcohorts (Table~\ref{tab:cohorts}): (a) $P_{cancer}$ (N=115) includes men who had confirmed prostate cancer and underwent radical prostatectomy, and (b) $P_{normal}$ (N=24), includes men without prostate cancer confirmed by negative MRI and negative biopsy. All men underwent multi-parametric Magnetic Resonance Imaging (MRI) prior to the biopsy or surgery. Our study included the T2-weighted (T2w) sequence and the Apparent Diffusion Coefficient (ADC) maps. For the patients in $P_{cancer}$, the resected tissue was sectioned in the same axial plane as the MR-sequence using a patient-specific 3D mold. The tissue was fixed and stained with Hematoxylin \& Eosin (H\&E)-staining and scanned at 20x magnification. The digitized histopathology images were then registered to the corresponding pre-operative MRI slices using the RAPSODI registration platform \citep{rusu_2020}, resulting in spatially aligned radiology and pathology images on a slice-level (Figure~\ref{fig:ex_prostate}). The Internal Review Board at Stanford University approved the retrospective study with waived patient consent. For more information about the cohort, and details on data processing and registration, please see~\citep{bhattacharya_2022, bhattacharya_2022a}. 


\paragraph{Labels and annotations}
Each patient in the $P_{cancer}$ cohort had annotations showing either normal tissue, indolent (Gleason Grade Group = 1) or aggressive (Gleason Grade Group $>$ 1) cancer, as well as the prostate gland outlined. On radiology, the annotations were outlined by a radiologist and confirmed by pathology findings. 
The annotations on the histopathology data were created automatically using a previously validated deep learning model~\citep{ryu_2019}, and processed using morphological processing and connected component analysis to ensure lesions were continuous in 3D. 
Lesions smaller than 250mm$^3$ were discarded as these are less likely to be seen on MRI and less likely to be clinically significant cancer~\citep{matoso_2019}. 
For more information about the annotation process, see \cite{bhattacharya_2022a}. 

The strategy used to create the simulated prostate data is described below in Section~\ref{sec:method}.

\subsubsection{Kidney Cohort}
For renal cell carcinoma, we used publicly available clinical data from the 2021 Kidney and Kidney Tumor Segmentation Challenge (KiTS21) and The Cancer Imaging Archive (TCIA): TCGA-KIRC. All patients included in this study had confirmed clear cell RCC, underwent contrast-enhanced CT scans, and, in the case of TCGA-KIRC, had corresponding diagnostic histopathology images from partial or radical nephrectomy. This resulted in 203 patients with CT from the KiTS21 dataset \citep{heller_2021}, and 174 patients with CT and histopathology from the TCGA-KIRC cohort \citep{akin_2016} (Table~\ref{tab:cohorts}). The tumor diameter (mean and standard deviation (SD)) was 5.01 (SD=3.11) cm for KiTS21 and 5.83 (SD=2.89) cm for TCGA-KIRC (see Table~\ref{tab:suppl_tumordim} for details). Figure~\ref{fig:ex_kidney} shows example data from a sample patient in the kidney cohort. 


\paragraph{Labels and annotations}
The KiTS21 data included manually segmented regions of the kidneys and tumors. For the TCGA-KIRC data, no manual segmentations were available. Automatic segmentations of the kidneys and tumors were extracted on CT using the previously validated and pre-trained MISCnn model~\citep{muller_2021}. These segmentations were visually inspected for quality assurance. On histopathology data, tissue was segmented from background using Otsu thresholding~\citep{otsu_1979}.

Each tumor was labeled as either indolent or aggressive using the pathology reports. Aggressive tumors were defined as having necrosis or a high tumor grade (Fuhrman~\citep{fuhrman_1982} or ISUP (International Society of Urological Pathology)~\citep{moch_2016} grade of 3 or 4). Indolent tumors were defined as those with low tumor grade (1-2) without necrosis ~\citep{bhindi_2018}. Tumor grade and necrosis were both included in the definition of aggressiveness, since both are considered important prognostic markers for clear cell RCC~\citep{fuhrman_1982, beddy_2014, tse_2021, zigeuner_2010}. Out of the total 372 tumors, 199 were labeled as aggressive. 19 low-grade (grade 2) cases had presence of necrosis (see Table~\ref{tab:suppl_necrosis} for details), thus considered aggressive despite low tumor grade. 
While clear cell RCC tumors may be heterogeneous, with both low- and high-grade areas, the lack of pixel-wise annotations restricted labels with finer granularity. Thus, the lesion segmentation masks were created by setting all values in the segmented lesion to the corresponding class. 

\paragraph{Pre-processing}
All CT volumes were cropped using a three-dimensional bounding box extracted from the lesion segmentations and resized to an in-plane dimension of 224x224 pixels. 
By using the intensity values inside the lesion, each cropped volume was Z-score normalized. Due to the variability of stain color in the public data, the histopathology images were stain normalized by the method from \cite{macenko_2009a}. For low-resolution feature extraction (described in Section~\ref{sec:method}), the data were downsampled to an image size of 224x224 pixels. 

%% file: sections/2b_method.tex
\begin{figure*}[!h]
    \centering
    \includegraphics[width=\linewidth]{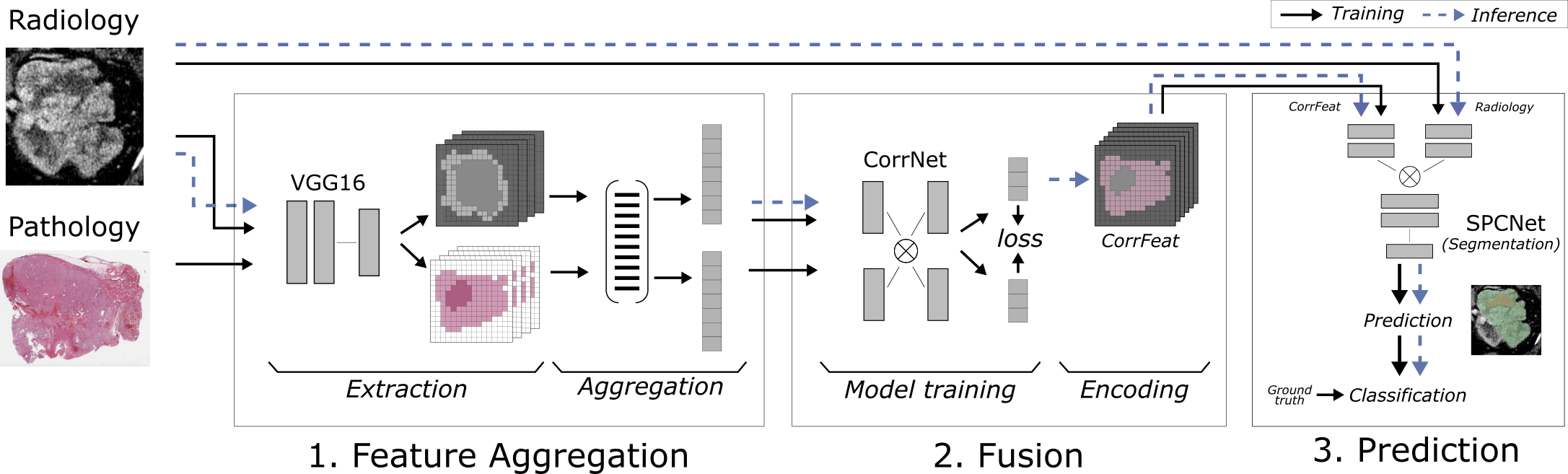}
    \caption{Summary of the proposed method \textit{CorrFABR}. The three-step approach consists of a feature aggregation step, a fusion step between the radiology and pathology data, and finally, incorporation of fused radiological image biomarkers to a prediction model, of which the output can be classification or segmentation map. During training, the radiology image(s), and the pathology image, are used in steps 1 and 2. The model trained in step 2 is applied to the radiology images, creating \textit{correlated features}, denoted \textit{CorrFeat}. These, together with the radiology data, are used as model input in step 3. During inference (blue dashed track), pathology images are not needed. }
    \label{fig:method}
\end{figure*}

\subsection{Method}
\label{sec:method}

This study aims to present a method for incorporating disease characteristics from histopathology images in to the radiology domain without requiring spatial alignment between the two domains. The overarching goal is to enable improved prediction performance using radiology images alone during inference. An overview of the proposed method, Correlated Feature Aggregation By Region (CorrFABR), is shown in Figure~\ref{fig:method}, which consists of three steps: 
\begin{enumerate}
    \item Feature extraction and aggregation from radiology and pathology images
    \item Fusion to learn a combination of radiology features that correlate with pathology features
    (denoted \textit{CorrFeat}) 
    \item Prediction to classify clear cell RCC tumors as aggressive or indolent, using either 
    the correlated features alone (\textit{CorrFeat}), or together with the radiology images.
\end{enumerate}

During training, unaligned radiology and pathology images originating from the same patient are used in Step 1 (Feature aggregation) and Step 2 (Fusion). For training of the model in the Step 3 (Prediction), as well as during inference, only radiology images are needed. Therefore, the model only relies on histopathology data during the training phase of two out of three steps, and never during inference. Below, each step of the method is described in more detail.  

\subsubsection{Input data}
During training, for each patient, image data from the radiology and pathology domains are used. In this work, we consider 3-dimensional radiology data (MRI or CT) where each slice containing the lesion is considered separately. Each slice is paired with a whole-slide image (WSI) from the histopathology domain. A pair formed by one radiology slice and one WSI are used as input to Step 1, Feature aggregation.


\subsubsection{Feature aggregation}
The first step of the method extracts feature vectors from the input images, which will be used as input to the fusion model in Step 2. Feature extraction was done using the first two layers of a pre-trained VGG16 model~\citep{bhattacharya_2022}. Image sizes of 224x224 pixels were used for radiology images, resulting in feature maps of shape [224, 224, 64], where 64 is the number of filter outputs from the second VGG16 layer~\citep{simonyan_2015}. For histopathology data, two approaches were evaluated. The first approach, denoted \textit{low-res}, assumes downsampled and resized data of size 224x224 pixels, giving feature maps of shape [224, 224, 64]. The second approach, denoted \textit{high-res}, extracts features at a higher resolution (between $0.008-0.032$ mm/pixel) in a patch-wise manner, with a patch size of 224x224 pixels. The number of patches per image depends on the width and height of the WSI. For each patch, the features are averaged, resulting in a 1x64 vector \textit{per patch}. 
Depending on the width ($w$) and height ($h$) of the input image, the feature maps are of shape [$w$/224, $h$/224, 64]. 

\begin{figure*}
    \centering
    \includegraphics[width=0.8\linewidth]{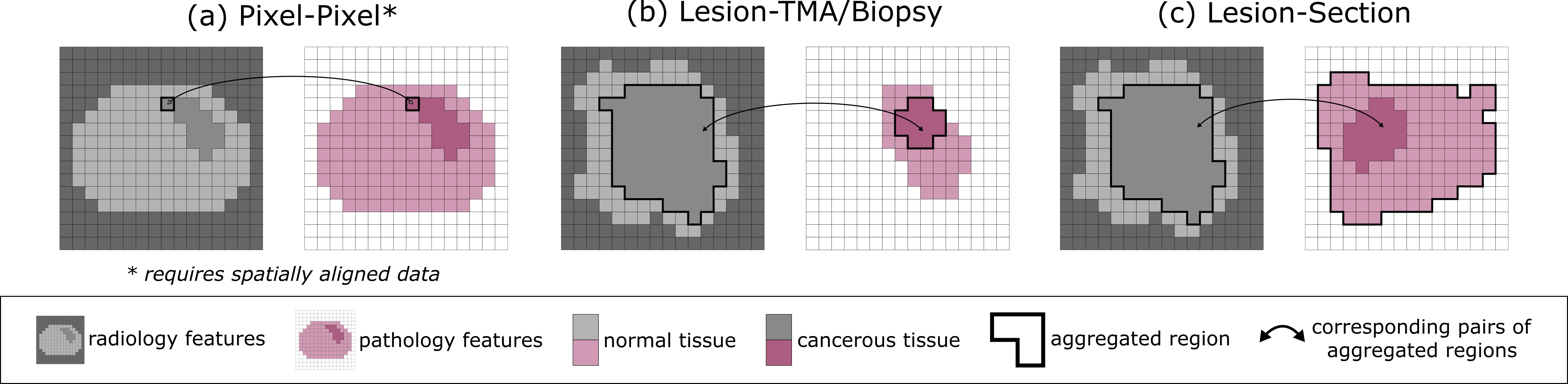}
    \caption{Schematic overview of the three different aggregation methods for creating paired feature vectors between radiology and pathology feature responses. For (b) and (c), pathology feature extraction can be done in both low- and high-resolution. (a) Pixel-pixel: feature vectors at each pixel position (requires spatially aligned data). 
    (b) Lesion-TMA/biopsy: smaller cutouts of selected tissue from the histopathology domain together with that lesion region from radiology. (c) Lesion-section: larger cutouts of tissue from the histopathology domain together with that lesion region in radiology. Figure best viewed in color.}
    \label{fig:feat_agg}
\end{figure*}

The extracted radiology and pathology features are then aggregated for fusion using three scenarios as below:

\paragraph{(a) Pixel-pixel fusion} (Figure \ref{fig:feat_agg} (a)): 
This feature aggregation applies to the scenario where the radiology and pathology images are spatially aligned, enabling pixel-level fusion, as used in \citep{bhattacharya_2022}. The aggregation of the features is done by flattening in the in-plane dimension, such that each corresponding pixel between the two domains, radiology and pathology, is considered an input-vector pair to the fusion model. Only \textit{low-res} aggregation is possible for this scenario, as it assumes the same resolution for both domains. Per patient, the set $S_{pixel} = \{S_{rad}, S_{pat}\}$ was formed, with sizes $S_{rad} = [224 \times 224, 64\times n]$ and $S_{pat} = [224 \times 224, 64]$, where $n$ denotes the number of input sequences available. In the case of the prostate cohort, $n=2$ (T2w and ADC), and for kidney $n=1$ (CT). Following \citep{bhattacharya_2022}, the number of input vectors was class balanced between cancer and non-cancer vectors, and limited to 1 million inputs per training set.

\paragraph{(b) Lesion-TMA/biopsy fusion} (Figure \ref{fig:feat_agg} (b)):
This feature aggregation is for the targeted biopsy scenario, where a small tissue microarray (TMA) is extracted from the targeted lesion. Typically, the TMA section is small, and may contain a mixture of cancerous tissue from the lesion and normal tissue from adjoining regions.  

In this scenario, the features were considered on a per-lesion basis instead of on a per-pixel basis. Feature values within the lesion regions were aggregated into one value in the in-plane dimension on a per-lesion basis in the radiology domain and per-TMA section basis in the pathology domain. For the extracted \textit{low-res} histopathology features, aggregation was done by simply taking the average. For the extracted \textit{high-res} histopathology features, the features were aggregated across the region by taking the 95th percentile of the intensity values. This approach ensures that local, high-resolution information is captured in the patches, and the aggregation over the region using the top percentile ensures that high, local, feature responses are maintained. 

In addition, normal tissue patches were extracted from randomly selected normal areas of $\approx$ 0.5x0.5 cm ($20x20$ pixels in \textit{low-res}). 


Per patient, the set $S_{biopsy} = \{S_{rad}, S_{pat}\}$ is now constructed by $j$ number of lesion regions and $k$ number of normal regions, with the size $S_{rad} = [(j+k), 64\times n]$ and $S_{pat} = [(j+k), 64]$.

\paragraph{(c) Lesion-section fusion} (Figure \ref{fig:feat_agg} (c)): This feature aggregation is for scenarios where the entire lesion has been removed via surgery.  The histopathology image consists of a large section of tissue containing cancerous tissue together with some adjacent normal tissue. In contrast to the TMA/biopsy scenario, the region is now larger, may be more heterogeneous, and may contain larger portions of normal tissue. Feature aggregation for this scenario was conducted similarly to lesion-TMA/biopsy scenario (b). Thus, per patient, we now have set $S_{section} = \{S_{rad}, S_{pat}\}$ constructed by $j$ number of lesion regions and $k$ number of normal regions, with the size $S_{rad} = [(j+k), 64\times n]$ and $S_{pat} = [(j+k), 64]$. 

\begin{figure}[b]
    \centering
    \includegraphics{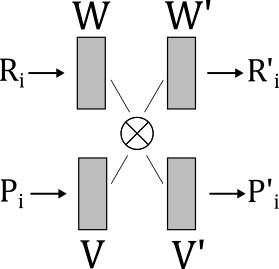}
    \caption{CorrNet model, with $R_i$ and $P_i$ as input representing one pair of feature vector. The input is processed through the single-layer encoder with weights $W$ and $V$, respectively. The correlation loss ($L_{corr}$) forces the resulting representations to be as correlated as possible. The output from the single-layer decoders, $W'$ and $V'$, reconstructs the input signal as closely as possible through the reconstruction loss ($L_{recon}$).}
    \label{fig:corrnet}
\end{figure}

\subsubsection{Fusion}
The fusion step enables learning radiology biomarkers that are correlated with pathology features. It consists of two parts: model training and encoding. The model considered in this work is a Correlational Neural Network (CorrNet)~\citep{chandar_2016}. CorrNet learns a common representation between two input vectors, such that they are maximally correlated in this lower dimensional latent space, while at the same time minimizing the reconstruction error between the output of the decoder and input. The reconstruction loss prevents collapse of the latent representation (e.g., all zeros).  

The CorrNet model architecture consists of one encoder-decoder pair for each input domain (see Figure~\ref{fig:corrnet}). Using one vector pair from $S$ (where $S$ is either $S_{pixel}$, $S_{biopsy}$ or $S_{section}$ from Step 1), the input vectors $R_i$ and $P_i$ are constructed, originating from radiology and pathology data respectively. The latent representations $h_R = f(W R_i + b)$ and $h_P = f(V P_i + b)$ are the output from the single-layer encoders of each domain, respectively. $f$ can be any linear or non-linear function, and $W \in  \mathbb{R}^{k \times 64 \times n}$, $V \in \mathbb{R}^{k \times 64}$ and $b \in \mathbb{R}^{k \times 1}$, where $k$ is the size of the latent representation and $n$ denotes the number of input sequences available. The joint hidden representation is given by

\begin{gather}
h(R_i, P_i) = f(W R_i + V P_i + b).
\end{gather}

The loss consists of two parts. The first part, denoted $L_{recon}$, minimizes the reconstruction of the between the input and the output of the decoder ($R'_i = g(W' h_R + b')$ and $P'_i = g(V' h_P + b')$, $g$ is any activation function.). The second part, denoted $L_{corr}$,  maximizes the correlation between $h_R$ and $h_P$.  The total loss is therefore: 

\begin{gather}
\centering
    L = L_{recon} + \lambda L_{corr},
\end{gather}

where $L_{recon}$ is given by

\begin{gather}
    L_{recon} = \sum^N_{i=1} \left(L(R_i, R'_i)) + L(P_i, P'_i)\right)
\end{gather}

and $N$ is the number of samples in one batch and  $L$ is the reconstruction error calculated using mean squared error. $L_{corr}$ is given by

\begin{gather}
    L_{corr} = \frac{\sum^N_{i=1} \left(h_R^i - \overline{h}_R \right)\left(h_P^i - \overline{h}_P \right) }{\sqrt{\sum^N_{i=1}\left(h_R^i - \overline{h}_R\right)^2 \left(h_P^i - \overline{h}_P\right)^2}}. 
\end{gather}

In this study, $\lambda = 2$ for all experiments, the dimension of the hidden space was set to $k=5$, and $f$ and $g$ are set as the identity function~\citep{bhattacharya_2020}.

Once the model is trained, only the encoder part of the radiology domain is used, applied to radiology data alone. This ensures that the method is usable for clinical inference, where the pathology data is unavailable. For inference, the input radiology features are flattened in the in-plane dimension. The output of the encoder, $h_R$ have size $[224 \times 224, 5]$, is reformatted as $[224, 224, 5]$, forming image-shaped features henceforth denoted \textit{CorrFeat}. These are the maximally correlated features between the radiology and pathology domain, constructed without pathology data and used as input to the third and final step, prediction.

\subsubsection{Prediction}

The SPCNet~\citep{seetharaman_2021} architecture was used as prediction model, taking both radiology input as well as correlated features (\textit{CorrFeat}) created from Step 2, outputting a segmentation mask. The architecture of SPCNet is based on the holistically nested edge detector (HED)~\citep{xie_2017}. SPCNet inputs three adjacent slices from the input radiology volume to predict the segmentation output of the center slice. One (1) or multiple sequences/inputs are inputted by separate branches consisting of convolutional layers, concatenated to form the segmentation output. In this work, we consider different scenarios depending on the available data. For prostate, the input radiology consists of two sequences, T2w and ADC. The SPCNet architecture has with this input two branches (denoted \textit{MRI}). In addition, a third input can be added for the CorrFeat, resulting in three branches (denoted \textit{MRI+CorrFeat}). For the kidney data, the input radiology consists of one CT volume, resulting in SPCNet architecture of one (1) branch (denoted \textit{CT}). When the CorrFeat are added, the number of branches is extended to two (denoted \textit{CT+CorrFeat}). In addition, for both prostate and kidney, the case where only the \textit{CorrFeat} are used as input is evaluated (one branch).

%% file: sections/2c_experiments.tex
\begin{figure*}
    
    \centering
    \begin{subfigure}[b]{0.15\textwidth}
        \centering
        \includegraphics[width=\textwidth]{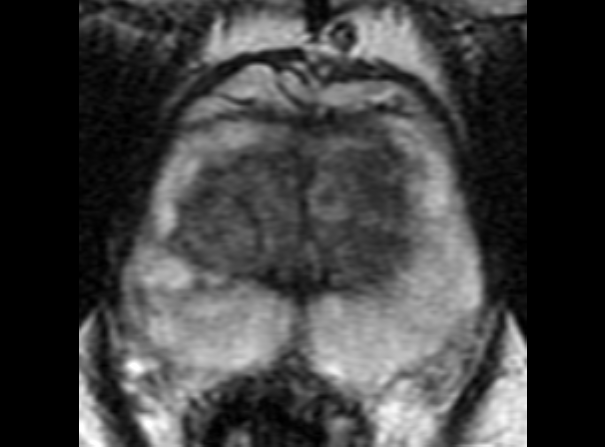}
    \end{subfigure}
    \begin{subfigure}[b]{0.15\textwidth}  
        \centering 
        \includegraphics[width=\textwidth]{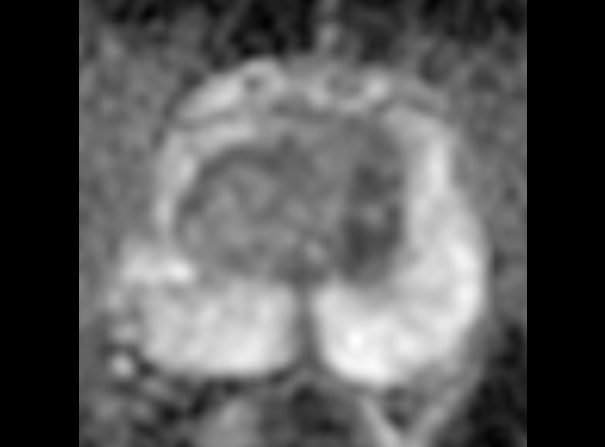}
    \end{subfigure}
    \begin{subfigure}[b]{0.15\textwidth}   
        \centering 
        \includegraphics[width=\textwidth]{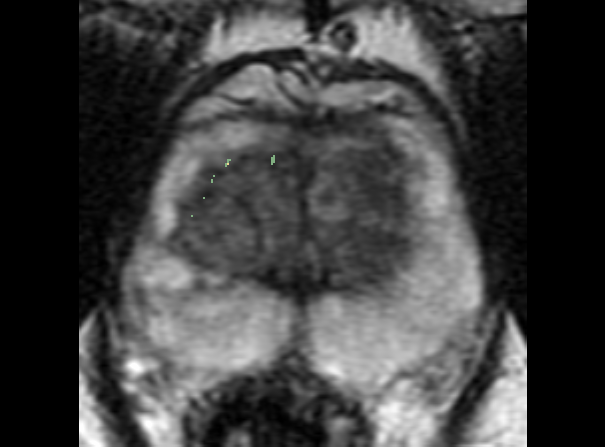}
    \end{subfigure}
    \begin{subfigure}[b]{0.15\textwidth}   
        \centering 
        \includegraphics[width=\textwidth]{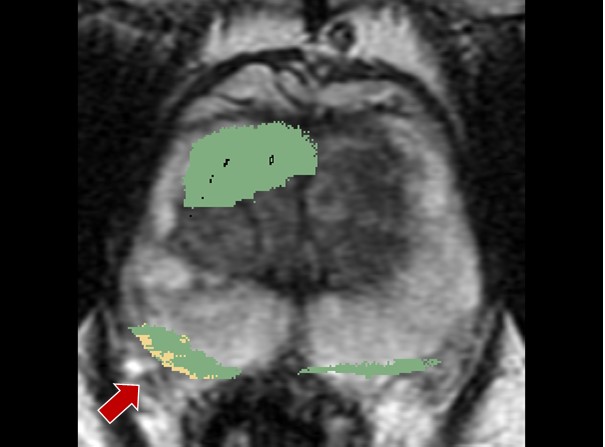}
    \end{subfigure}
    \begin{subfigure}[b]{0.15\textwidth}   
        \centering 
        \includegraphics[width=\textwidth]{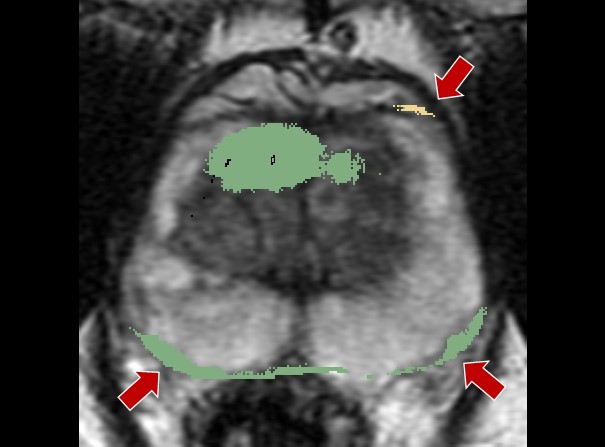}
    \end{subfigure}
    \begin{subfigure}[b]{0.15\textwidth}   
        \centering 
        \includegraphics[width=\textwidth]{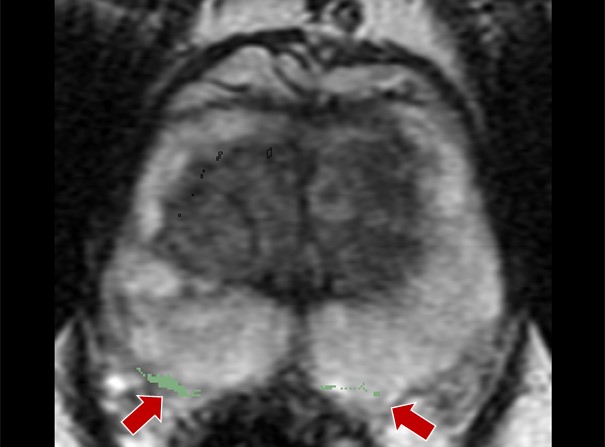}
    \end{subfigure}
    \centering
    \begin{subfigure}[b]{0.15\textwidth}
        \centering
        \includegraphics[width=\textwidth]{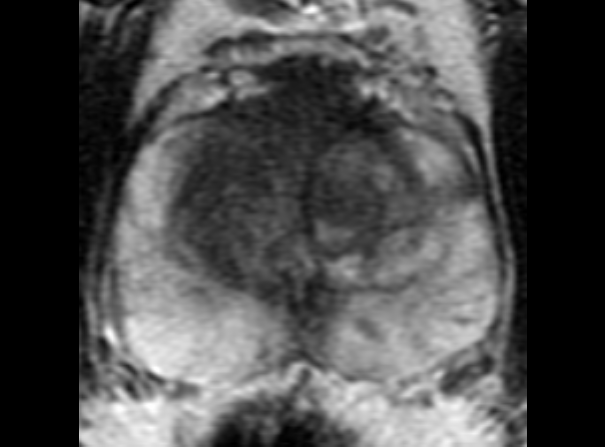}
    \end{subfigure}
    \begin{subfigure}[b]{0.15\textwidth}  
        \centering 
        \includegraphics[width=\textwidth]{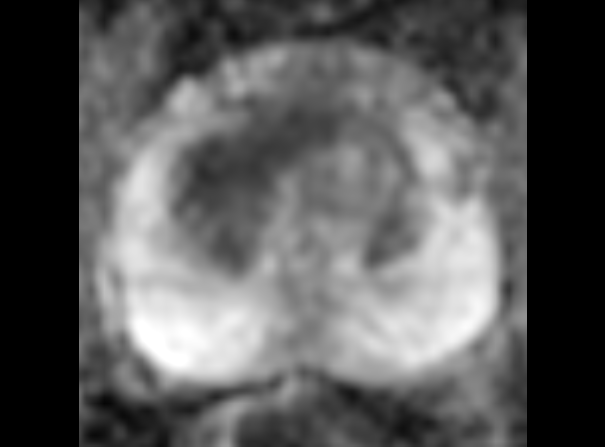}
    \end{subfigure}
    \begin{subfigure}[b]{0.15\textwidth}   
        \centering 
        \includegraphics[width=\textwidth]{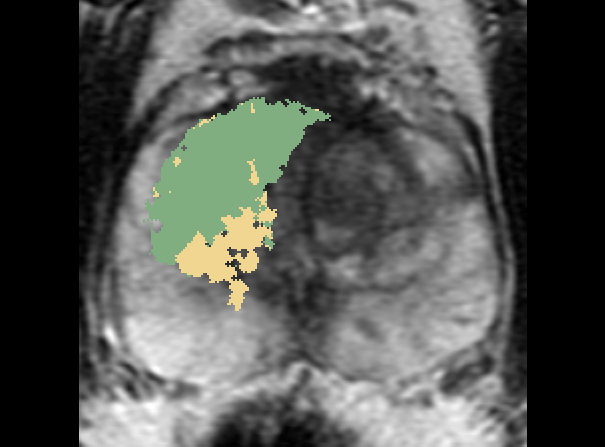}
    \end{subfigure}
    \begin{subfigure}[b]{0.15\textwidth}   
        \centering 
        \includegraphics[width=\textwidth]{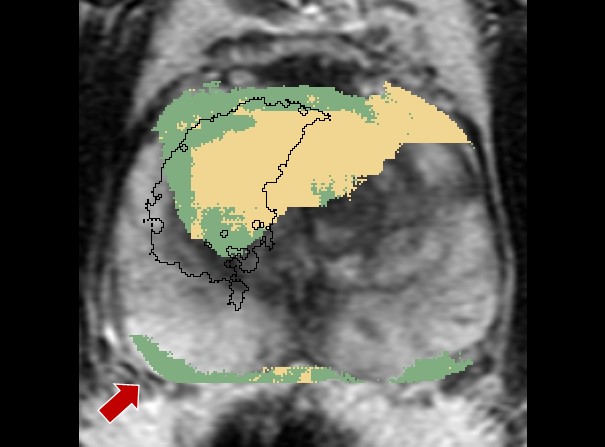}
    \end{subfigure}
    \begin{subfigure}[b]{0.15\textwidth}   
        \centering 
        \includegraphics[width=\textwidth]{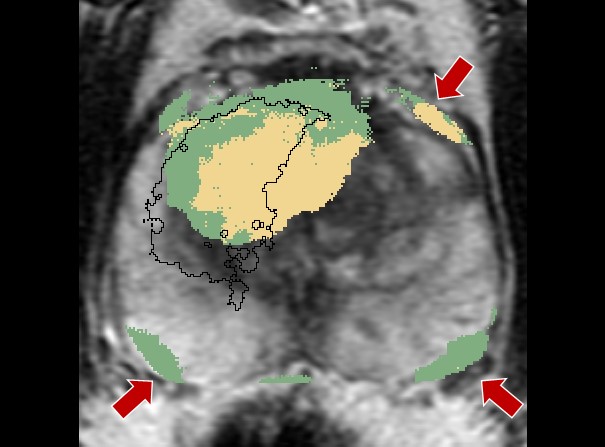}
    \end{subfigure}
    \begin{subfigure}[b]{0.15\textwidth}   
        \centering 
        \includegraphics[width=\textwidth]{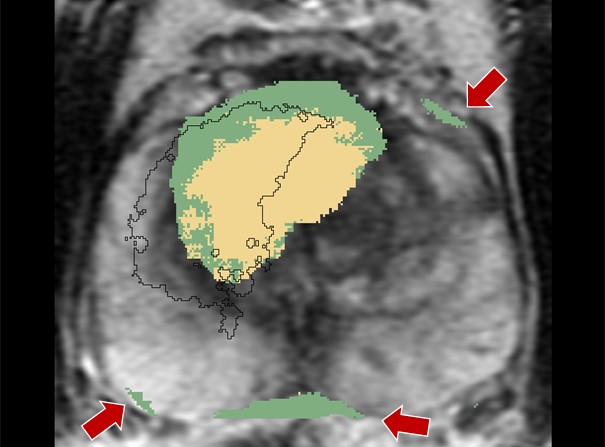}
    \end{subfigure}
    \centering
    \begin{subfigure}[b]{0.15\textwidth}
        \centering
        \includegraphics[width=\textwidth]{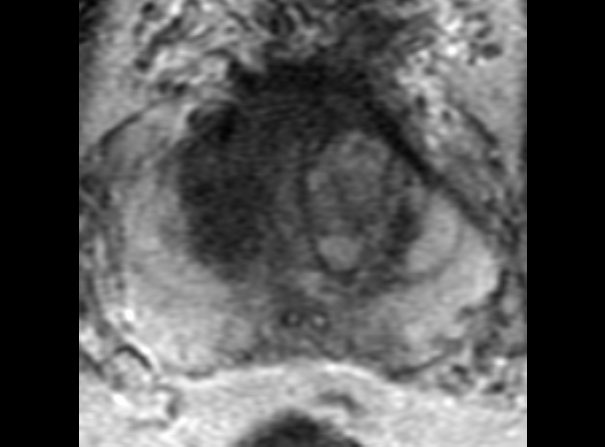}
        \caption{T2w}
        \label{fig:prostate_t2}
    \end{subfigure}
    \begin{subfigure}[b]{0.15\textwidth}  
        \centering 
        \includegraphics[width=\textwidth]{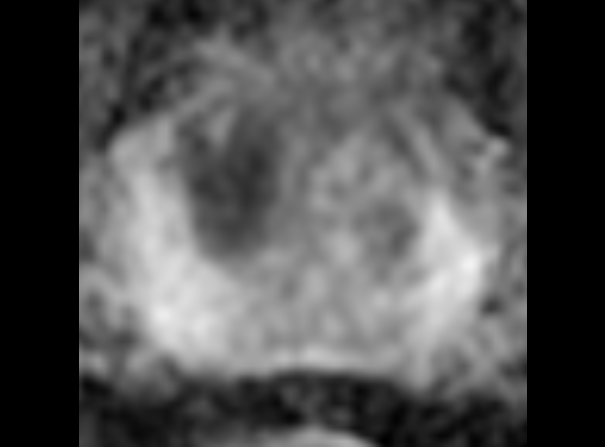}
        \caption{ADC}
        \label{fig:prostate_adc}
    \end{subfigure}
    \begin{subfigure}[b]{0.15\textwidth}   
        \centering 
        \includegraphics[width=\textwidth]{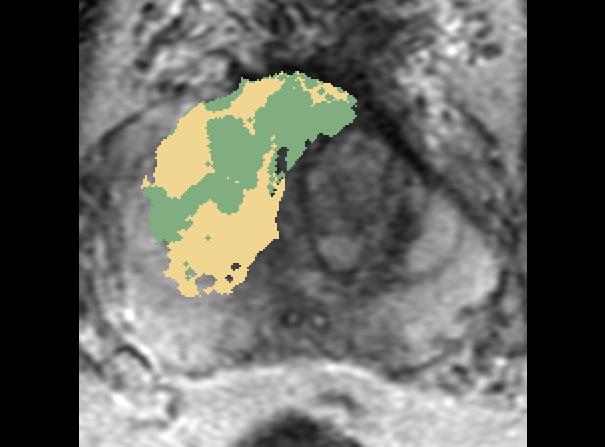}
        \caption{Ground truth}
        \label{fig:prostate_gt}
    \end{subfigure}
    \begin{subfigure}[b]{0.15\textwidth}   
        \centering 
        \includegraphics[width=\textwidth]{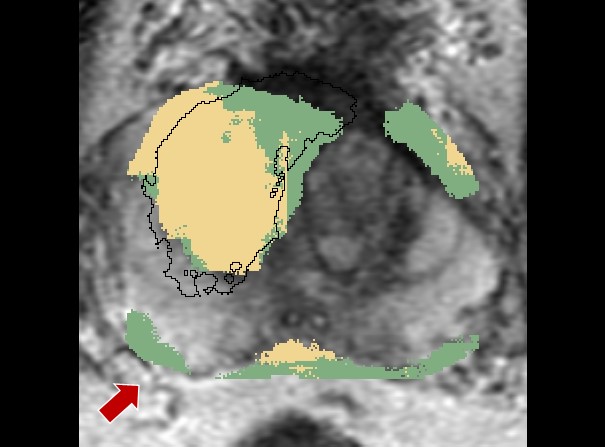}
        \caption{SPCNet}
        \label{fig:prostate_spc}
    \end{subfigure}
    \begin{subfigure}[b]{0.15\textwidth}   
        \centering 
        \includegraphics[width=\textwidth]{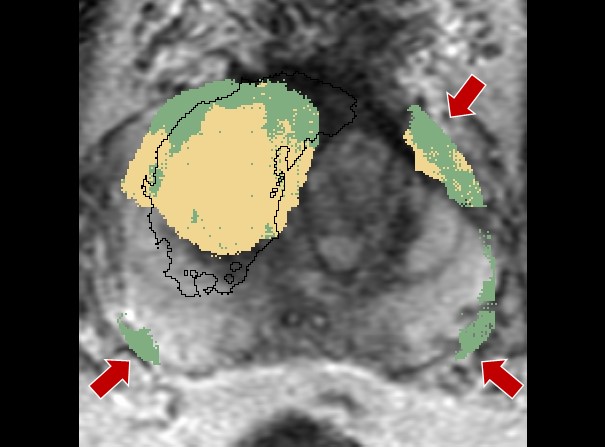}
        \caption{CorrSigNIA}
        \label{fig:prostate_cor}
    \end{subfigure}
    \begin{subfigure}[b]{0.15\textwidth}   
        \centering 
        \includegraphics[width=\textwidth]{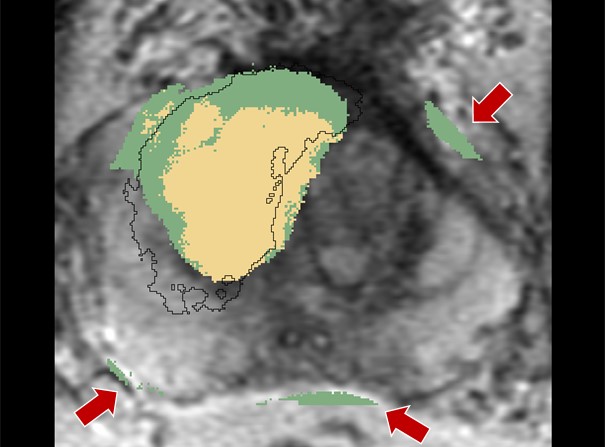}
        \caption{CorrFABR}
        \label{fig:prostate_sec}
    \end{subfigure}
    
    \caption{Three slices from one patient from the prostate test cohort with an aggressive tumor in the anterior and central part of the prostate. The three left-most panels show T2w, ADC, and T2w with ground truth segmentations overlayed (indolent regions in green, aggressive in yellow). The right-most panels show prediction results for (d) SPCNet (no CorrFeat), (e) CorrSigNIA (pixel-pixel CorrFeat), and (f) CorrFABR (lesion-section CorrFeat), with ground truth overlayed (black outline). Arrows show examples of false positive predictions. }
    \label{fig:prostate_e1}
\end{figure*}

\begin{table}[]
\centering
\caption{Number of unique cases in training and test splits for different tasks. }
\label{tab:traintest}
\resizebox{\linewidth}{!}{%
\begin{tabular}{@{}lcccccc@{}}
\toprule
Task & \multicolumn{3}{c}{\textit{E1: Prostate Pixel-level agg.}} &  \multicolumn{3}{c}{\textit{E2: Kidney Tumor-level agg.}}  \\ \cmidrule(lr){2-4} \cmidrule(lr){5-7}
 Cohort & $P_{cancer}$ & $P_{normal}$ & Total &  Indolent & Aggressive & Total   \\ \midrule
Train & 75 & 24 & 99 &  136 & 162 & 298 \\
Test & 40 & 0 & 40 &  37 & 37 & 74  \\ \bottomrule
\end{tabular}%
}
\end{table}

\begin{table*}[]
\centering
\caption{Lesion-wise results for aggressive prostate cancer detection on MRI (Experiment E1). Qualitative results for SPCNet, CorrSigNIA, and CorrFABR with lesion-section aggregation from high-resolution pathology are visualized in Figure~\ref{fig:prostate_e1}.}
\label{tab:e1_cs_latest}
\resizebox{\textwidth}{!}{%
\begin{tabular}{@{}lcclcccccc@{}}
\toprule
Method & Pathology Resolution & Spatial Alignment & Feat. Aggregation & Input Rad. & Input \textit{CorrFeat} & ROC-AUC & Dice & Sensitivity & Specificity \\ \midrule
SPCNet & N/A & N/A & None & \checkmark & - & 0.86 ± 0.28 & 0.36 ± 0.19 & 0.86 ± 0.34 & 0.48 ± 0.35 \\
CorrSigNIA & Low & \checkmark & Pixel-pixel & \checkmark & \checkmark & 0.87 ± 0.29 & 0.37 ± 0.21 & 0.86 ± 0.34 & 0.60 ± 0.37 \\
CorrFABR & Low & - & Lesion-TMA/biopsy & \checkmark & \checkmark & 0.89 ± 0.27 & 0.36 ± 0.21 & 0.82 ± 0.37 & 0.53 ± 0.36 \\
CorrFABR (\textit{visualized}) & Low & - & Lesion-section & \checkmark & \checkmark & 0.90 ± 0.22 & 0.36 ± 0.19 & 0.94 ± 0.22 & 0.51 ± 0.37 \\ 
CorrFABR & High & - & Lesion-section & \checkmark & \checkmark & 0.87 ± 0.31 & 0.36 ± 0.22 & 0.86 ± 0.34 & 0.59 ± 0.37 \\ 
CorrFABR  & High & - & Lesion-section & - & \checkmark & 0.84 ± 0.30 & 0.32 ± 0.21 & 0.82 ± 0.38 & 0.49 ± 0.34 \\ \bottomrule
\end{tabular}%
}
\end{table*}

\section{Experiments and results}
\label{sec:experiments}
We use two experimental setups to train and evaluate CorrFABR. In the first experiment, data from prostate cancer patients with existing spatial correspondences between radiology and histopathology images are used. The alignment is synthetically removed to simulate more common data-availability scenarios originating from targeted biopsy or surgery. We train CorrFABR to learn correlated MRI features in this simulated prostate scenario, with the goal of selectively identifying aggressive and indolent prostate cancer on MRI. In the second experiment, we train and evaluate CorrFABR to classify clear cell RCC lesions as aggressive and indolent using CT images from publicly available data. In both scenarios, there is no spatial alignment or slice-to-slice correspondence between the radiology and pathology images.

\subsection{Experiment 1 (E1): Prostate}

In this experiment, we synthetically modify the unique spatially aligned data to simulate ``real-world'' data for the prostate, where (a) pathology images are acquired either through biopsy or after surgery, and (b) exact registration or spatial correspondence between radiology and pathology images is unknown. 
The goal of this experiment was not to improve performance over previous methods for prostate lesion segmentation, but to enable development and assessment of radiology-pathology fusion models in simulated prostate scenarios when spatial alignment is unavailable. Another goal was to increase understanding of how to modify the fusion approaches for renal cell carcinoma, where only a rough estimate of the location of the pathology image is known. This is in contrast to previously presented radiology-pathology fusion methods~\citep{bhattacharya_2020, bhattacharya_2022}, which relied on both in-plane spatial alignment, as well as slice-to-slice correspondence between the radiology and pathology images. 


\subsubsection{Dataset}
Two different aggregation scenarios were considered (see Section~\ref{sec:method} and Figure~\ref{fig:feat_agg}): lesion-TMA/biopsy (simulating targeted biopsy) and lesion-section (simulating surgery).  
In the radiology domain, for both scenarios, lesion-level T2w and ADC features were extracted and aggregated from regions within the lesion outline on a per-slice basis, resulting in one aggregated feature vector per lesion of size $1 \times 128$. Features were also aggregated from normal regions by randomly sampling the same number of normal patches as lesion vectors from a single patient. 

In the pathology domain, smaller and larger regions were extracted using lesion annotations to simulate real-world biopsy and surgery scenarios. The selected histopathology sections were randomly offset beyond the lesion outlines to include some random fraction of normal tissue in addition to a minimum threshold of cancerous tissue. To further simulate the biopsy and surgery scenario, \textit{high-res} features were extracted from pathology (as described in Section~\ref{sec:method}). 

Feature vectors from the histopathology images are paired with the corresponding lesion in radiology, but from a randomly selected slice. Feature vectors from normal regions are similarly paired with vectors from normal regions, but from randomly selected slices and locations. 

As an intermediate step, the method was evaluated in the case where spatial alignment was kept, but features were aggregated on a per-region basis. Results for this experiment is shown in the Supplementary Material.

\subsubsection{Training and test set split}
For the prostate cohorts, 99 patients from $P_{cancer}$ and $P_{normal}$ cohorts were used for training, and 40 cases from $P_{cancer}$ were used for independent testing (Table \ref{tab:traintest}). As only cases from $P_{cancer}$ had corresponding histopathology, these cases were used in Step 1 and 2 of training (feature extraction, aggregation, and training of the fusion model, described in Section~\ref{sec:method}), but all cases were used for Step 3 (prediction) of training. 

The models were trained in a 5-fold cross-validation setting, by splitting the 99 cases in training into five equally sized groups, such that each group had 20 (or 19) unique cases. For each group, a model was trained on all cases not included, and the remaining cases were used to validate the model. 
The folds were kept fixed during all three steps of the method. The final segmentation outputs of the five models were averaged to form one final prediction segmentation, from which the evaluation metrics were calculated.

\subsubsection{Training details and evaluation}
Two models were considered baseline: the SPCNet model with only radiology images as input~\citep{seetharaman_2021}, and the prior method from \cite{bhattacharya_2022}, \textit{CorrSigNIA}, which used MRI + \textit{CorrFeat} as input, where \textit{CorrFeat} was learned using pixel-pixel feature aggregation and fusion. 

For Step 2, the CorrNet models were trained for 1000 epochs, with learning rate $0.5 \cdot 10^{-4}$ and batch size 50. For Step 3, the SPCNet models were trained for maximally 100 epochs, a learning rate of 0.001, and batch size of 8. The learning rate was reduced by 0.1 if the validation loss had not been reduced in 10 epochs. Early stopping was used with a patience of 20 epochs, and the final model was used for evaluation. 

The prediction model was trained to selectively identify normal tissue, indolent cancer, and aggressive cancer on the entire prostate. 
The models were evaluated in detecting and localizing aggressive prostate cancer on MRI using a sextant-based approach used in previous studies~\citep{seetharaman_2021, bhattacharya_2022, bhattacharya_2022a}. Models using MRI-only (baseline), MRI + \textit{CorrFeat}, as well as \textit{CorrFeat} only were evaluated. ROC-AUC, Dice, sensitivity, and specificity are reported for detection of lesion-level aggressive cancer (see below).

\subsubsection{Results}

Qualitatively, both CorrSigNIA and CorrFABR resulted in fewer false-positive regions identified compared to the baseline SPCNet method that did not incorporate any correlated features (Figure~\ref{fig:prostate_e1}). Furthermore, there is less over-segmentation using the CorrFABR method, especially notable in the top slice. Please see Figure~\ref{fig:suppl_prostate_e1} for qualitative results in another patient.


Quantitatively, CorrFABR demonstrated better or similar performance compared to the baseline CorrSigNIA approach, despite lacking spatial alignment between the radiology and pathology images (Table~\ref{tab:e1_cs_latest}). Comparing the high-resolution and low-resolution approaches for feature extraction from the histopathology data, the Dice score is nearly identical. This suggests that the extraction of histopathology features at high resolution is not needed. The ROC-AUC and sensitivity were higher for the low-resolution case, but at the cost to a lower specificity. CorrFABR with low-resolution and lesion-section aggregation resulted in higher ROC-AUC, sensitivity and specificity, and similar Dice compared to baseline. Using only \textit{CorrFeat} as input, without MRI gave a slightly lower result in terms of ROC-AUC, Dice, and sensitivity (bottom row), indicating the need for the structural information present in the MRI data. Overall, the results show the benefit of using correlated feature learning in the real-world scenario where perfect registration is not available between radiology and pathology images.

\begin{table*}[]
\centering
\caption{Lesion-wise results for clear cell RCC tumor aggressiveness classification (Experiment E2). Quantitative results for SPCNet and CorrFABR with lesion-section aggregation from high resolution pathology (\textit{CorrFeat} only) are visualized in Figure~\ref{fig:kidney_e3}.}
\label{tab:e2_results}
\resizebox{\textwidth}{!}{%
\begin{tabular}{@{}lcclcccccc@{}}
\toprule
Method & Pathology Resolution & Spatial Alignment & Feat. Aggregation & Input Rad. & Input \textit{CorrFeat} & ROC-AUC & F1-score & Sensitivity & Specificity \\ \midrule
VGG16 & N/A & N/A & None & \checkmark & - & \textbf{0.78 ± 0.03} & 0.68 ± 0.04 & 0.62 ± 0.16 & 0.76 ± 0.13 \\
SPCNet & N/A & N/A & None & \checkmark & - & 0.72 ± 0.05 & 0.66 ± 0.02 & 0.49 ± 0.07 & \textbf{0.85 ± 0.04} \\
CorrFABR & Low & - & Lesion-section & \checkmark & \checkmark & 0.74 ± 0.04 & 0.68 ± 0.06 & 0.63 ± 0.10 & 0.75 ± 0.13 \\
CorrFABR & High & - & Lesion-section & \checkmark & \checkmark & 0.75 ± 0.03 & 0.67 ± 0.04 & 0.58 ± 0.12 & 0.77 ± 0.17 \\
CorrFABR (\textit{visualized}) & High & - & Lesion-section & - & \checkmark & \textbf{0.78 ± 0.02} & \textbf{0.73 ± 0.03} & \textbf{0.64 ± 0.08} & 0.83 ± 0.06 \\ \bottomrule
\end{tabular}%
}
\end{table*}

\begin{figure}[!h]
\centering
    \begin{subfigure}{\linewidth}
    \centering
        \includegraphics[width=\linewidth, trim={0 0.2cm 0 0.3cm},clip]{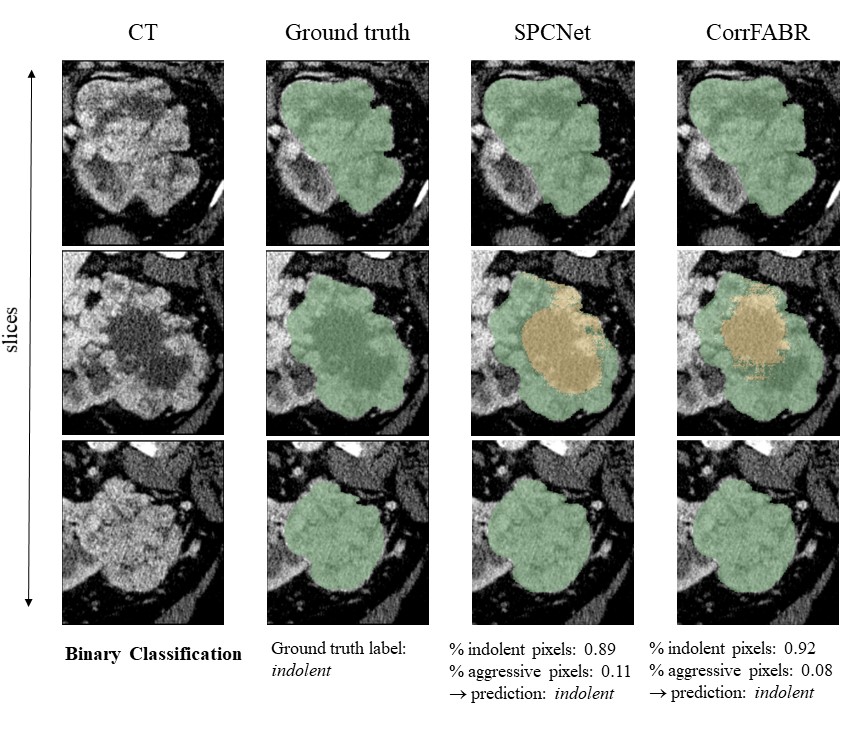}
    \end{subfigure}
    \begin{subfigure}{\linewidth}
    \centering
        \includegraphics[width=\linewidth, trim={0 0.2cm 0 0.3cm},clip]{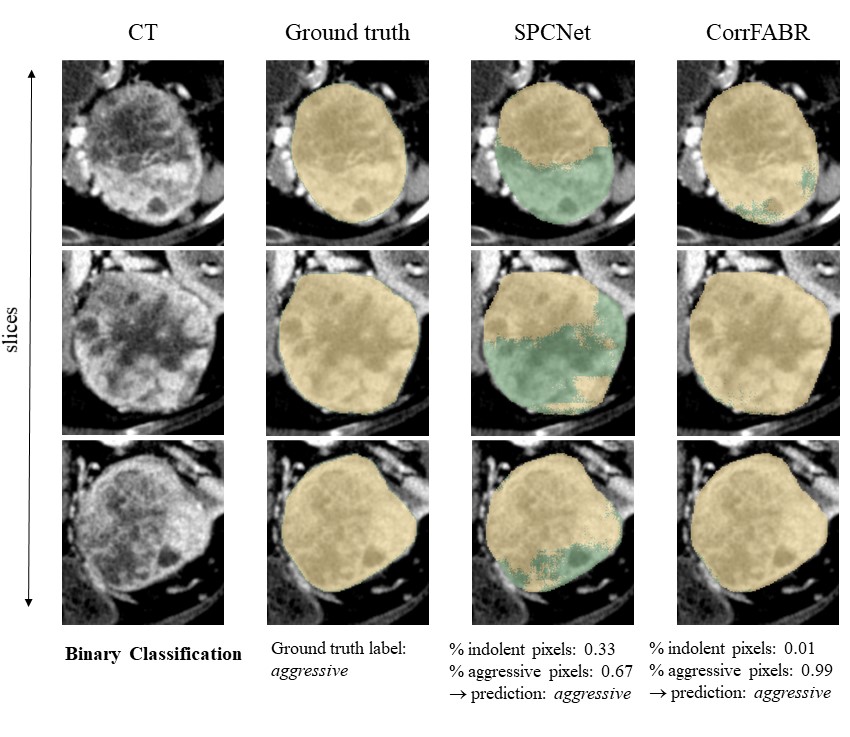}
    \end{subfigure}
    \begin{subfigure}{\linewidth}
    \centering
        \includegraphics[width=\linewidth, trim={0 0.2cm 0 0.3cm},clip]{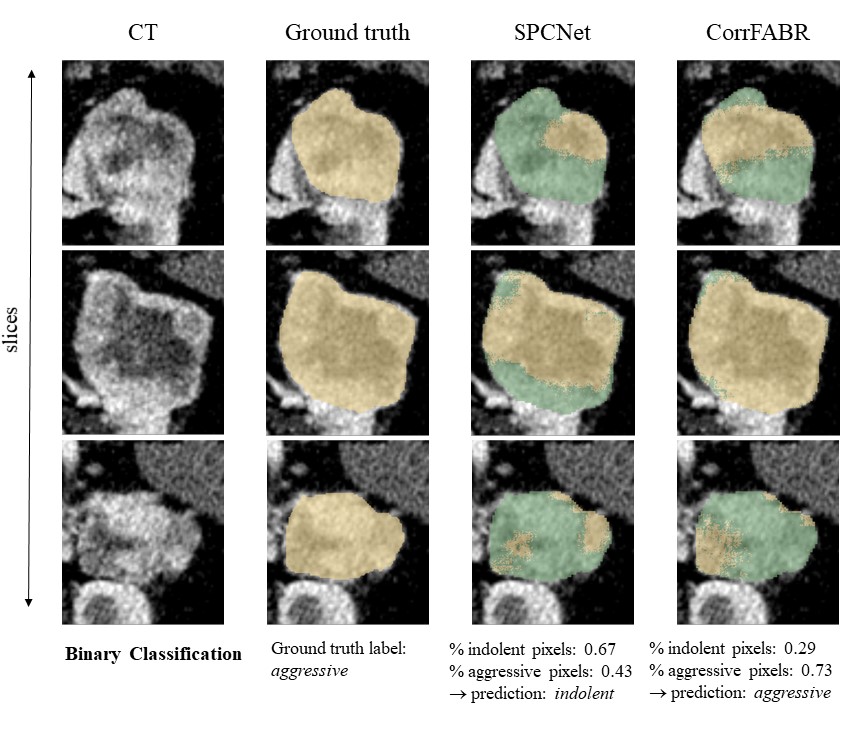}
    \end{subfigure}
    \caption{Results for clear cell classification, showing three slices from three different cases. Green denote indolent pixels, yellow aggressive. The segmentation values are converted to binary classification by majority vote. Best viewed in color.}
    \label{fig:kidney_e3}
\end{figure}

\subsection{Experiment 2 (E2): clear cell renal cell carcinoma aggressiveness prediction}
\label{sec:ex2}
In this experiment, the approach described in E1 was adapted to train models on the kidney cohort for clear cell RCC tumor aggressiveness prediction. This cohort naturally lacks both pixel-level and slice-level correspondence, and the histopathology data consist of a single slice per patient, with a section of the surgically removed tissue used for diagnosis (see Figure~\ref{fig:ex_kidney}). 

\subsubsection{Dataset}
The lesion-section feature aggregation was used for this task, as the histopathology images were from surgically removed sections of the lesion with no spatial alignment with the radiology images. Moreover, the histopathology images lacked cancer annotations. Feature extraction of the pathology features was evaluated both for the \textit{low-res} and the \textit{high-res} approach.

For each slice in the 3-dimensional CT volume, the features were extracted and aggregated from the annotated lesion region, similar to methods used for the prostate. For the histopathology data, only one whole-slide pathology image was available per patient. Due to lack of annotations, no normal regions were extracted. The feature vector extracted and aggregated from the entire tissue section (from either low- or high-resolution images) was paired with each of the slices in CT volume.

\subsubsection{Training and test set split}
A total of 378 cases were included from KiTS21 and TCGA-KIRC. A test set of 74 cases with balanced distribution of aggressive and indolent clear cell RCC was randomly selected from KiTS21, to ensure that the test set included cases of high segmentation quality. The remaining cases not included in the test set formed a 5-fold cross-validation setup (Table~\ref{tab:traintest}). As only TCGA-KIRC included CT and histopathology data, this cohort was used to train in steps 1 and 2, while training cases from KiTS21 were included in Step 3. The evaluation metrics are reported as mean and standard deviation between models and folds on a per-lesion level. 

\subsubsection{Training details and evaluation}
For incorporation of the \textit{CorrFeat}, the SPCNet model was used, evaluated on both CT + \textit{CorrFeat}, and \textit{CorrFeat} only. A five-fold cross-validation approach was used with the same training hyperparameters as used for experiment E1, for both the CorrNet model and the SPCNet model. As a baseline prediction model, a VGG16 model was evaluated using CT data only. Using transfer learning from pre-trained ImageNet weights, the model was trained similarly with early stopping, a learning rate of 0.001, and batch size of 8. To avoid overfitting, only the last layer was fine-tuned. While model weights are available from the prostate Experiment E1, they were not used as part of Experiment E2 due to the different nature of the input data and pathology appearance. 

For the kidney task, the goal was to discriminate between indolent or aggressive cancer on a per-lesion basis. As the SPCNet model was trained for a segmentation endpoint, the output segmentation volumes were converted to binary class labels on a per-patient level by setting the predicted class as the one with the largest number of predicted pixels. The reported metrics are ROC-AUC, F1-score, sensitivity, and specificity (see below). 

\subsubsection{Results}
This experiment evaluates the method on data where there is only a patient-level correspondence between the radiology and pathology data and no spatial alignment between the images. 

From qualitative inspection of the segmentation outputs, we see that the incorporation of pathology-correlated radiology image biomarkers (proposed method) helps to better discriminate the aggressive and indolent tissue characteristics on CT images by labeling more pixels as the correct class (Figure~\ref{fig:kidney_e3}). Especially in one of the aggressive cases (middle), the CorrFABR method correctly classifies large portions in the center of the tumor, which the SPCNet model misclassified as indolent regions. 

Quantitatively, comparing baseline methods of VGG (pre-trained on ImageNet) and SPCNet that neither incorporate correlated data from histopathology gave similar performance as to adding the correlated features together with CT data (Table~\ref{tab:e2_results}). However, when the correlated features were used stand-alone using high-resolution histopathology features (bottom row), the F1-score improved from $0.68 \pm 0.04$ and $0.66 \pm 0.02$ respectively to $0.73 \pm 0.03$. Furthermore, both low- and high-resolution approaches gave similar results, indicating the feasibility of using either approach for correlated feature training by region.

%% file: sections/4_conclusion.tex
\section{Discussion}
\label{sec:discussion}
We presented a novel method, CorrFABR (Correlated Feature Aggregation By Region), for incorporation of disease characteristics from histopathology images to help in radiologic assessment without the need for spatially aligned radiology and pathology images. By extending previous approaches~\citep{bhattacharya_2020, bhattacharya_2022}, the proposed method creates correlated features from aggregated regions from each image domain, removing the need for pixel-to-pixel aligned data. 
Starting from spatially aligned radiology and pathology prostate images, we synthetically remove the pixel- and slice correspondence to mimic the more common data-availability scenario. Despite this, it is still possible to extract valuable correlated information that improves prostate cancer detection over MRI features alone (Table~\ref{tab:e1_cs_latest}), even better than when pixel-to-pixel correlation are used. This benefit can be attributed to the fact that even when pixel-to-pixel correspondences are assigned, registration errors exist. In the case of the prostate, they are estimated to be between 2-3mm \citep{rusu_2020}. By doing region-based correlation learning, the effect of the registration error is reduced.  This motivated us to apply the approach to the more challenging real-world data of clear cell renal cell carcinoma (RCC).

Accordingly, the method was evaluated on public CT data for tumor aggressiveness classification of clear cell RCC. By using both high tumor grade (grade 3 or 4) as well as the presence of necrosis as definition of aggressiveness, we incorporated two important prognostic factors~\citep{minardi_2005, hotker_2016, rabjerg_2021}. 

Previous methods have been presented for tumor-grade classification only, without reference to necrosis ~\citep{ding_2018, kocak_2019a, bektas_2019, lin_ct-based_2019, sun2019prediction, shu_clear_2019, cui_2020, lin_2020, xv_2021, demirjian_2021}. Therefore, numerical comparison of results is challenging, as the target endpoint is different.  This is further amplified by evaluation on internal data only and/or the use of different/multiple input data sequences (unenhanced, varying delay times of enhanced images, or multiple phases).


Comparing the results from the first experiment of simulated prostate data with the second experiment of real-world kidney data, we see that the method generalized to (a) a different disease beyond the prostate, (b) a different radiology image (CT vs. MRI), (c) data without radiology-pathology registration, and (d) data with increased noise (heterogeneous data from different scanners and acquisition protocols) than the internal prostate data. 
For clear cell RCC, using radiology image biomarkers that correlate with pathology (proposed method) as stand-alone input gave higher performance than in combination with CT images, which was not the case for the prostate. We hypothesize that for clear cell classification, the correlated features already capture the discriminating features. Adding CT increases the model complexity, degrading its performance and increases the risk of overfitting. 

Our study has a few noteworthy limitations. First, the method is evaluated on clear cell RCC, and does not include other renal cancer subtypes. 
Secondly, by using public data, CT images taken at different post-contrast phases were included. This introduces noise in the signal to the model. A more homogeneous dataset could likely improve the models' performance, but potentially at the cost of generalizability across phases. 
Secondly, kidney tumors may be heterogeneous in that they have regions of both low- and high-grade areas, with more or less necrotic tissue. This granularity of tissue type was not reflected in the segmentation labels, as only a per-tumor label (indolent or aggressive) was available. However, this heterogeneity can be reflected at a global feature level by extracting high-dimensional features from entire sections (that may include normal, low-grade, high-grade and/or necrotic regions). Using radiology image biomarkers that correlate with pathology on a per-section level may therefore capture the variance in tissue types. This hypothesis is supported by the results from the prostate cohort, where this granularity of segmentation labels exists and where per-section feature aggregation gave similar or better performance compared to per-pixel.
Finally, two different grading systems were used for the different cohorts, Fuhrman and ISUP. \cite{dagher_2017} showed that ISUP is a better prognostic marker than Fuhrman when using all four grades (1-4). However, when grouping the grades into low and high, only 10 cases out of 279 differed between the grading systems (all 10 cases were Fuhrman grade 3 but ISUP grade 2). The impact of using two grading systems in this study is therefore considered to be low, and where a larger variance may come from high inter-reader variability within each tumor grade system~\citep{rabjerg_2021}. 



Our study is the first to bring pathology information into the radiology domain for clear cell RCC characterization in the clinically relevant scenario when radiology and pathology image are not registered and pathology images are not needed during inference. Our promising results demonstrate the applicability of our approach to other diseases with unaligned radiology and pathology images. Future work will focus on increasing the cohort size, using data from multiple institutions to improve generalization, as well as the performance of our model.

\section{Conclusion}

In this paper, we present CorrFABR, a method of creating radiological image biomarkers by incorporating  histopathology image features into the radiology domain without the need for spatially aligned data. During inference, the radiological image biomarkers can be extracted without dependence on histopathology images. 
We show that this approach improves performance in differentiating between indolent and aggressive clear cell renal cell carcinoma on CT compared to CT images alone. 
This independence of histopathology data during inference and removal of the need for spatially aligned data increases the clinical utility and relevance of the method for renal cancer characterization. 

%% file: sections/supplementary_material.tex
\newpage\section*{Supplementary Material}
\beginsupplement

\subsection*{S1. Prostate experiment, region-wise aggregation}
\label{sec:suppl_regionagg}

In Table~\ref{tab:suppl_e1_results}, results are included in the intermediate step where slice and in-plane correspondence between the radiology and pathology data exists, but where features are aggregated on a per-region basis (both lesion-TMA/biopsy and lesion-section). Results showing where this correspondence has been removed is also presented as comparison, as well as in Table~\ref{tab:e1_cs_latest}.

\begin{figure*}
    \centering
    \begin{subfigure}[b]{0.15\textwidth}
        \centering
        \includegraphics[width=\textwidth]{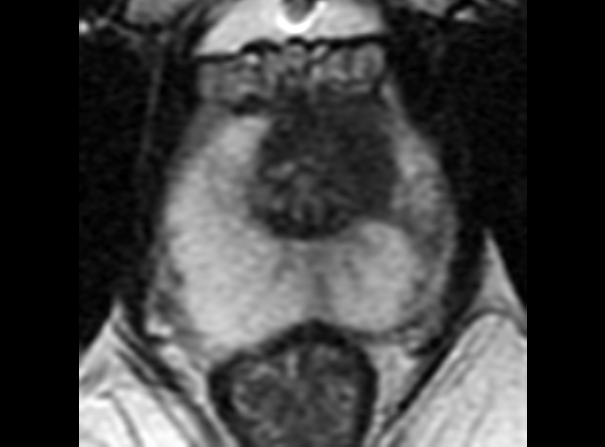}
    \end{subfigure}
    \begin{subfigure}[b]{0.15\textwidth}  
        \centering 
        \includegraphics[width=\textwidth]{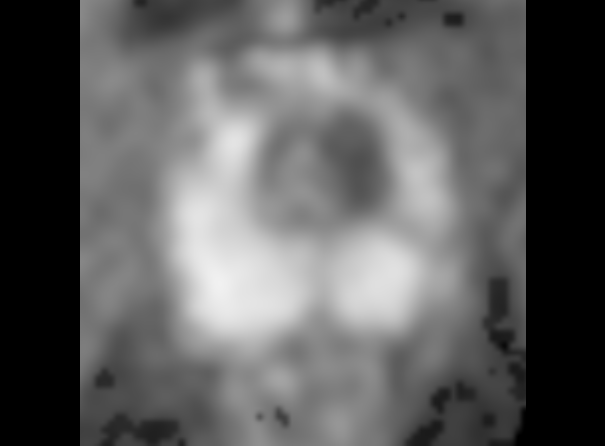}
    \end{subfigure}
    \begin{subfigure}[b]{0.15\textwidth}   
        \centering 
        \includegraphics[width=\textwidth]{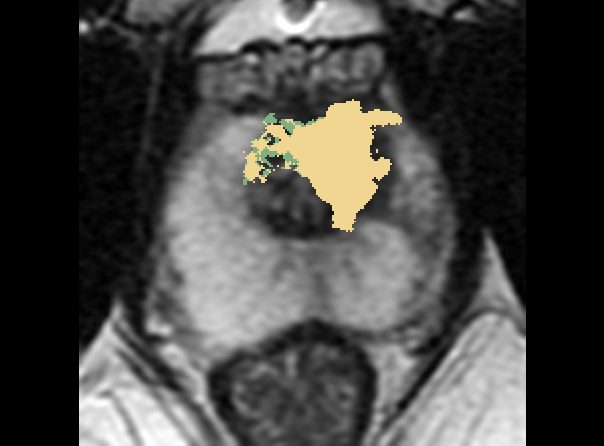}
    \end{subfigure}
    \begin{subfigure}[b]{0.15\textwidth}   
        \centering 
        \includegraphics[width=\textwidth]{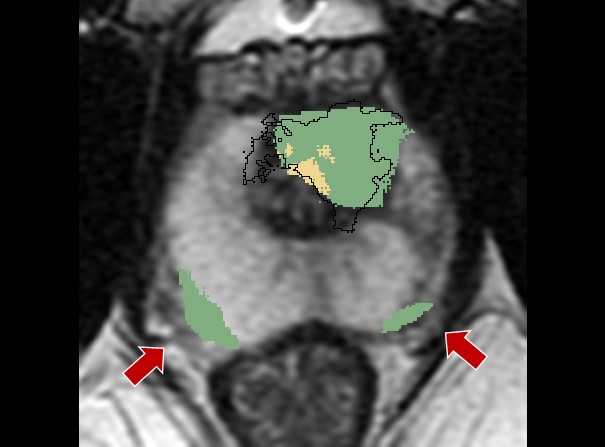}
    \end{subfigure}
    \begin{subfigure}[b]{0.15\textwidth}   
        \centering 
        \includegraphics[width=\textwidth]{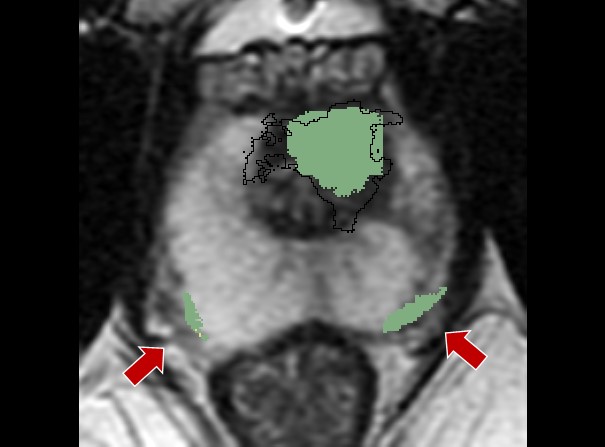}
    \end{subfigure}
    \begin{subfigure}[b]{0.15\textwidth}   
        \centering 
        \includegraphics[width=\textwidth]{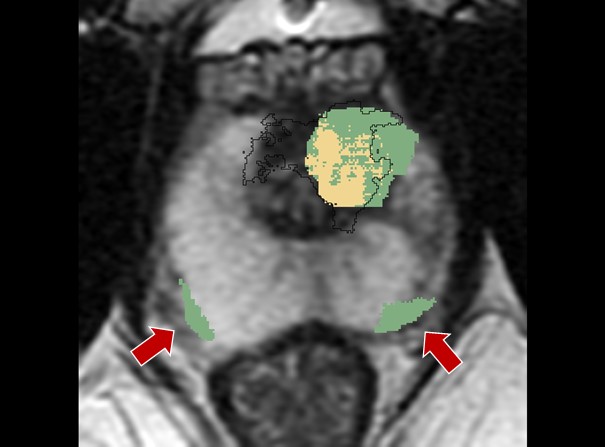}
    \end{subfigure}
    \centering
    \begin{subfigure}[b]{0.15\textwidth}
        \centering
        \includegraphics[width=\textwidth]{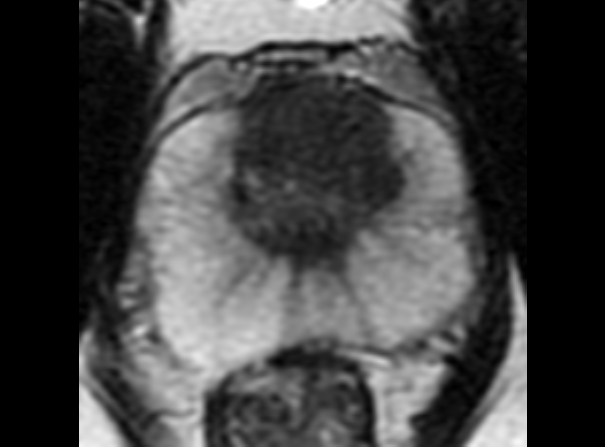}
    \end{subfigure}
    \begin{subfigure}[b]{0.15\textwidth}  
        \centering 
        \includegraphics[width=\textwidth]{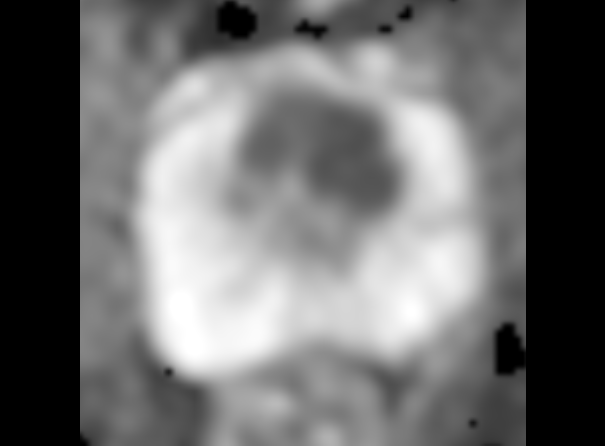}
    \end{subfigure}
    \begin{subfigure}[b]{0.15\textwidth}   
        \centering 
        \includegraphics[width=\textwidth]{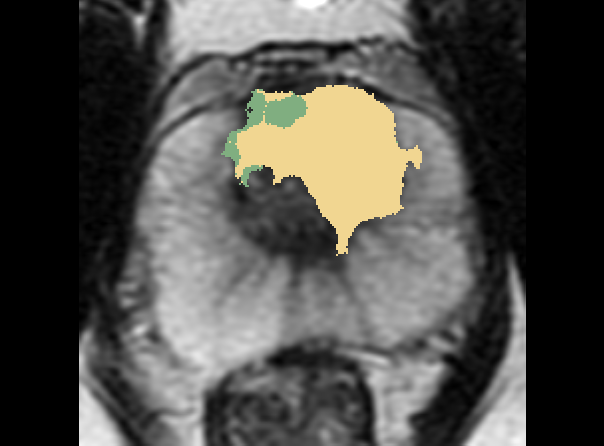}
    \end{subfigure}
    \begin{subfigure}[b]{0.15\textwidth}   
        \centering 
        \includegraphics[width=\textwidth]{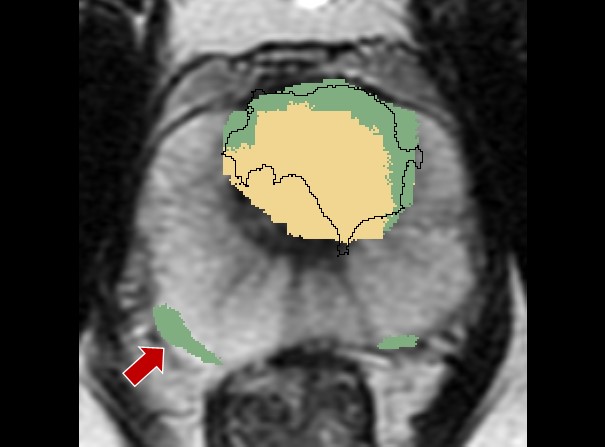}
    \end{subfigure}
    \begin{subfigure}[b]{0.15\textwidth}   
        \centering 
        \includegraphics[width=\textwidth]{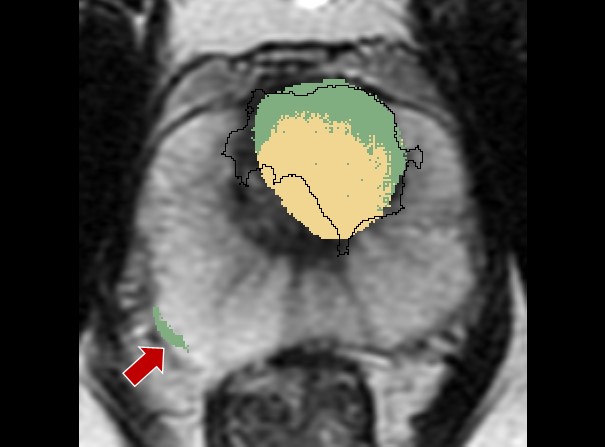}
    \end{subfigure}
    \begin{subfigure}[b]{0.15\textwidth}   
        \centering 
    \includegraphics[width=\textwidth]{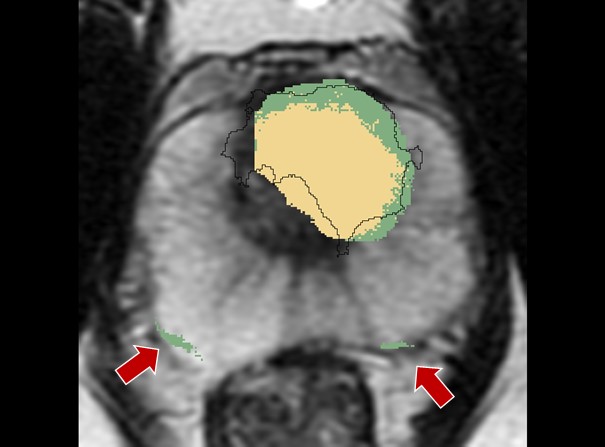}
    \end{subfigure}
    \centering
    \begin{subfigure}[b]{0.15\textwidth}
        \centering
        \includegraphics[width=\textwidth]{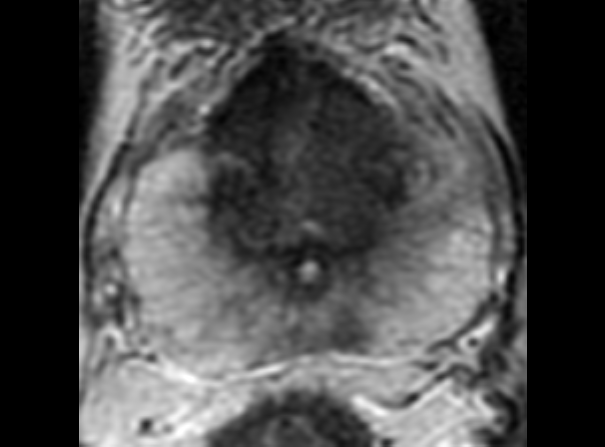}
        \caption{T2w}
        \label{fig:prostate_t2}
    \end{subfigure}
    \begin{subfigure}[b]{0.15\textwidth}  
        \centering 
        \includegraphics[width=\textwidth]{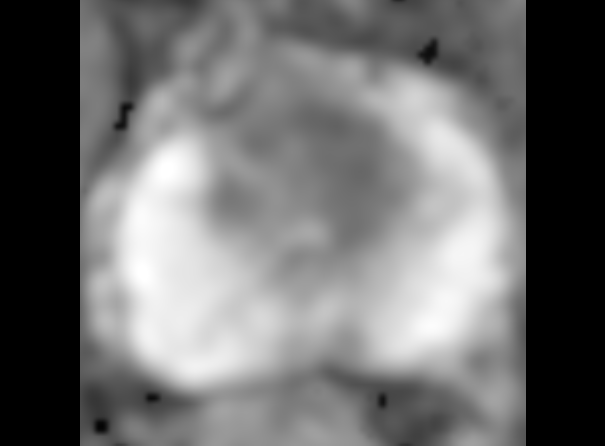}
        \caption{ADC}
        \label{fig:prostate_adc}
    \end{subfigure}
    \begin{subfigure}[b]{0.15\textwidth}   
        \centering 
        \includegraphics[width=\textwidth]{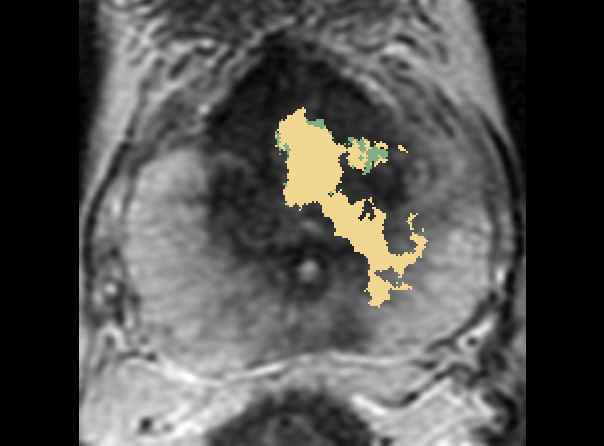}
        \caption{Ground truth}
        \label{fig:prostate_gt}
    \end{subfigure}
    \begin{subfigure}[b]{0.15\textwidth}   
        \centering 
        \includegraphics[width=\textwidth]{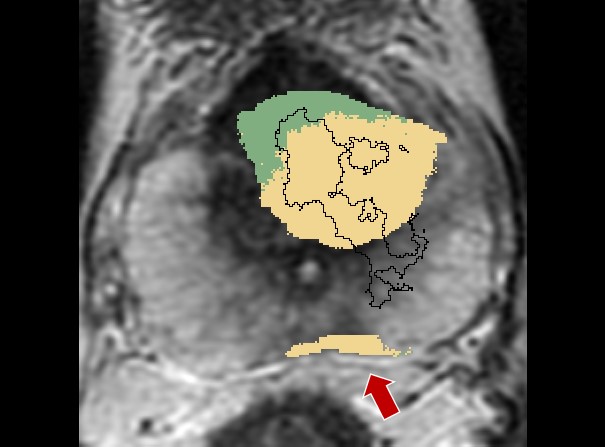}
        \caption{SPCNet}
        \label{fig:prostate_spc}
    \end{subfigure}
    \begin{subfigure}[b]{0.15\textwidth}   
        \centering 
        \includegraphics[width=\textwidth]{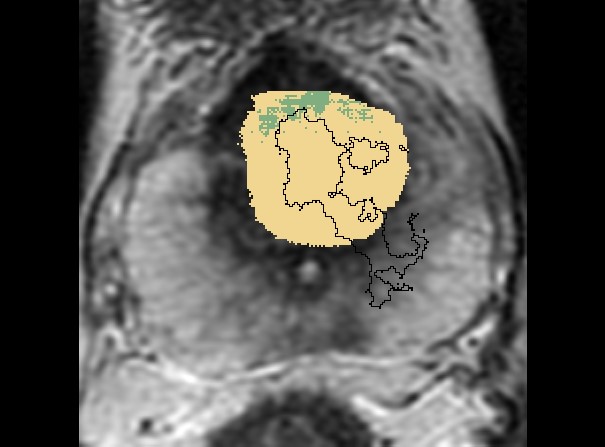}
        \caption{CorrSigNIA}
        \label{fig:prostate_cor}
    \end{subfigure}
    \begin{subfigure}[b]{0.15\textwidth}   
        \centering 
        \includegraphics[width=\textwidth]{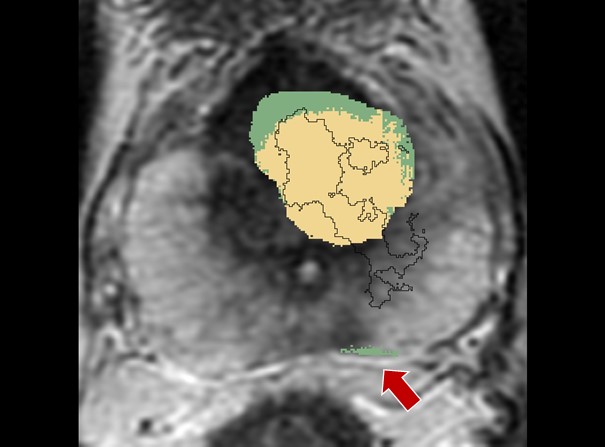}
        \caption{CorrFABR}
        \label{fig:prostate_sec}
    \end{subfigure}
    
    \caption{Example of a patient from the prostate test cohort, showing three slices. The three left-most panels show T2w, ADC and T2w with ground truth segmentations overlayed. The right-most panels show prediction results for (d) SPCNet, (e) CorrSigNIA (pixel-pixel CorrFeat), and (f) CorrFABR (lesion-section CorrFeat), with ground truth overlayed (black outline). Arrows shows exampled of false positive predictions. }
    \label{fig:suppl_prostate_e1}
\end{figure*}

\begin{table*}[]
\centering
\caption{Lesion-wise results for clinically significant prostate cancer, showing results with and without slice correspondence (E1).}
\label{tab:suppl_e1_results}
\resizebox{\textwidth}{!}{%
\begin{tabular}{@{}lcclccccc@{}}
\toprule
Method & Resolution & Slice Corr. & Feat. Aggregation & Input & ROC-AUC & Dice & Sensitivity & Specificity \\ \midrule
SPCNet & N/A & N/A & None & MRI & 0.86 ± 0.28 & 0.36 ± 0.19 & 0.86 ± 0.34 & 0.48 ± 0.35 \\
CorrSigNIA & Low & \checkmark & Pixel-pixel & MRI + CorrFeat & 0.87 ± 0.29 & 0.37 ± 0.21 & 0.86 ± 0.34 & 0.60 ± 0.37 \\
CorrFABR & Low & \checkmark & Lesion-TMA/biopsy & MRI + CorrFeat & 0.87 ± 0.26 & 0.35 ± 0.23 & 0.82 ± 0.38 & 0.64 ± 0.39 \\
CorrFABR & Low & - & Lesion-TMA/biopsy & MRI + CorrFeat & 0.89 ± 0.27 & 0.36 ± 0.21 & 0.82 ± 0.37 & 0.53 ± 0.36 \\
CorrFABR & Low & \checkmark & Lesion-section & MRI + CorrFeat & \textbf{0.90 ± 0.25} & \textbf{0.38 ± 0.21} & 0.90 ± 0.29 & 0.64 ± 0.36 \\
CorrFABR & Low & - & Lesion-section & MRI + CorrFeat & \textbf{0.90 ± 0.22} & 0.36 ± 0.19 & \textbf{0.94 ± 0.22} & 0.51 ± 0.37 \\ 
CorrFABR & Low & \checkmark & Lesion-section & CorrFeat & 0.87 ± 0.31 & \textbf{0.38 ± 0.23} & 0.86 ± 0.34 & \textbf{0.76 ± 0.31} \\ 
CorrFABR & High & - & Lesion-section & MRI + CorrFeat & 0.87 ± 0.31 & 0.36 ± 0.22 & 0.86 ± 0.34 & 0.59 ± 0.37 \\
CorrFABR & High & - & Lesion-section & CorrFeat & 0.84 ± 0.30 & 0.32 ± 0.21 & 0.82 ± 0.38 & 0.49 ± 0.34 \\ \bottomrule
\end{tabular}%
}
\end{table*}

\begin{table}[!h]
\centering
\caption{Tumor diameter in cm. For KiTS21, radiologic measurements are given. For TCGA-KIRC, approximations of diameter based on calculated volume from automatic segmentations are given. TCGA-KIRC has no Grade 1 cases.}
\label{tab:suppl_tumordim}
\resizebox{\linewidth}{!}{%
\begin{tabular}{@{}lcccc@{}}
\toprule
Tumor Grade & \multicolumn{1}{l}{Grade 1} & \multicolumn{1}{l}{Grade 2} & \multicolumn{1}{l}{Grade 3} & \multicolumn{1}{l}{Grade 4} \\ \midrule
KiTS21 & 3.24 ± 1.17 & 4.07 ± 2.31 & 6.27 ± 3.67 & 8.49 ± 2.70 \\
TCGA-KIRC & - & 4.96 ± 2.60 & 5.82 ± 2.72 & 8.00 ± 3.01 \\ \bottomrule
\end{tabular}%
}
\end{table}

\begin{table}[h!]
\centering
\caption{Number of cases with/without necrosis per tumor grade. Bold numbers indicate cases defined as aggressive (with necrosis and/or tumor grade 3 or 4). }
\label{tab:suppl_necrosis}
\begin{tabular}{@{}lccc@{}}
\toprule
 & Without Necrosis & With Necrosis & Total \\ \midrule
Grade 1 & 24 & \textbf{0} & 24 \\
Grade 2 & 149 & \textbf{19} & 168 \\
Grade 3 & \textbf{107} & \textbf{27} & 134 \\
Grade 4 & \textbf{23} & \textbf{23} & 46 \\ \midrule
Total & 303 & 69 & 372 \\ \bottomrule
\end{tabular}
\end{table}